\documentclass[twocolumn]{aastex631}
\usepackage[utf8]{inputenc}
\usepackage{amsmath,amsxtra,amssymb,latexsym,amscd,amsthm}
\usepackage[mathscr]{eucal}
\usepackage{mathrsfs}
\usepackage{makeidx}
\usepackage{soul}
\usepackage{graphicx}
\usepackage{subfigure}
\usepackage{booktabs}
\usepackage{multirow}
\usepackage{threeparttable}
\usepackage{hyperref}

\newcommand{\mbf}[1]{\mathbf{#1}}
\newcommand{\mbs}[1]{\boldsymbol{#1}}
\newcommand{\mrs}[1]{\mathscr{#1}}
\newcommand{\ovl}[1]{\overline{#1}}
\newcommand{\tx}[1]{\text{#1}}
\newcommand{\pt}{\partial}
\newcommand{\txbf}[1]{\textbf{#1}}

\shorttitle{A Multi-Fluid Dust Module in Athena++}
\shortauthors{Huang \& Bai}
\graphicspath{{./}{figures/}}

\begin{document}

\title{A Multi-Fluid Dust Module in Athena++: Algorithms and Numerical Tests}

\author[0000-0002-7575-3176]{Pinghui Huang}
\affiliation{Institute for Advanced Study, Tsinghua University, Beijing 100084, People’s Republic of China}

\author[0000-0001-6906-9549]{Xue-Ning Bai}
\affiliation{Institute for Advanced Study, Tsinghua University, Beijing 100084, People’s Republic of China}
\affiliation{Center for Astrophysics, Department of Astronomy, Tsinghua University, Beijing 100084, People’s Republic of China}

\email{phhuang@mail.tsinghua.edu.cn; xbai@tsinghua.edu.cn}

\begin{abstract}
  We describe the algorithm, implementation and numerical tests of a multifluid dust module in the Athena++ magnetohydrodynamic (MHD) code. The module can accommodate an arbitrary number of dust species interacting with the gas via aerodynamic drag (characterized by the stopping time), with a number of numerical solvers. In particular, we describe two second-order accurate, two-stage, fully-implicit solvers that are stable in stiff regimes including short stopping time and high dust mass loading, and they are paired with the second-order explicit van-Leer and Runge-Kutta gas dynamics solvers in Athena++, respectively. Moreover, we formulate a consistent treatment of dust concentration diffusion with dust back-reaction, which incorporates momentum diffusion and ensures Galilean invariance. The new formulation and stiff drag solvers are implemented to be compatible with most existing features of Athena++, including different coordinate systems, mesh refinement, shearing-box and orbital advection. We present a large suite of test problems, including the streaming instability in linear and nonlinear regimes, as well as local and global setting, which demonstrate that the code achieves the desired performance. This module will be particularly useful for studies of dust dynamics and planet formation in protoplanetary disks.
\end{abstract}


\section{Introduction}~\label{sec:intro}

Protoplanetary disks (PPDs) are composed of gas and dust. Although sharing only about 1\% in mass, dust represents the fundamental building blocks of planets, and it is primarily the thermal radiation from dust that makes PPDs observable in continuum emission from infrared to millimeter wavelengths. Dust is coupled with gas via aerodynamic drag, characterized by the stopping time. Dust particles of small sizes have small stopping time and are strongly coupled to the gas, while larger dust particles are more loosely coupled, and hence do not necessarily trace the gas. This fact is not only important many processes of planet formation, but also crucial for interpreting disk observations.

The initial stage of planet formation involves dust growth and transport, both of which are sensitive to disk structure and level of turbulence~\citep{OrmelCuzzi2007, BirnstielDullemond2010}. In particular, disk turbulence leads to dust diffusion~\citep{Cuzzi1993,YoudinLithwick2007,CarballidoCuzzi2010,ZhuStone2015}, which determines the thickness of the dust layer in the vertical direction, as well as the mixing in the radial direction. Additional ``psudo-diffusion” can result from complex radial gas flow structures due to, e.g., wind-driven accretion~\citep{HuBai2021}. Upon growing to larger sizes, back-reaction from dust-to-gas leads to dust clumping due to the streaming instability~\cite[SI,][]{GoodmanPindor2000,YoudinGoodman2005}, and subsequently planetesimal formation~\citep{JohansenOishi2007}. While there has been a large number of further studies~\cite[e.g.,][]{Bai2010dynamics,Bai2010pressure,CarreraJohansen2015,SimonArmitage2017,YangJohansen2017,LiYoudin2021}, it is less clear how the SI interplay with more realistic gas dynamics~\cite[see][]{JohansenKlahr2011,SchaferJohansen2020,XuBai2022}. Finally, instead of planetesimal accretion, the growth of planetary cores by pebble accretion has been identified to be more efficient towards higher core mass~\citep{OrmelKlahr2010Pebble,LambrechtsJohansen2012Pebble}. The efficiency of pebble accretion again depends on disk structure and level of turbulence~\cite[e.g.][]{Morbidelli2015,XuBai2017}, and back-reaction from dust-to-gas may destabilize the feeding zone~\citep{FuLi2014,PierensLin2019,YangZhu2020,HuangLiIsella2020,HsiehLin2020,SurvilleMayer2020}, which requires careful study considering realistic gas dynamics in 3D.

Over the past decade, thanks to the advent of the Atacama Large Millimeter/submillimeter Array (ALMA), as well as high-contrast imaging techniques equipped in ground-based telescopes, the dramatically improved resolution and sensitivity have led to the discovery of disk substructures prevalent in PPDs, particularly in the form of rings and gaps, as well as various forms of asymmetries~\cite[see][for a review]{Andrews2020}. These features are commonly interpreted as a consequence of planet-disk interaction, which can open gaps~\citep{BaeZhu2017,DongLi2017,DongLi2018}, create vortices~\citep{vanderMarel2013,ZhuStone2014,Flock2015}, drive spirals~\citep{DongRafikov2011I,DongRafikov2011II,BaeZhu2018I,BaeZhu2018II}, etc. At millimeter/sub-millimeter wavelength, the observed substructures reflect the distribution of mm-sized dust particles, which likely substantially amplify substructures in the gaseous disk because these particles are not strongly tied to gas and tend to drift towards pressure maxima~\citep{Whipple1972,Weidenschilling1977Aerodynamics}. Alternatively, a number of non-planet mechanisms have been identified which lead to substructure formation, such as processes involving snow lines~\citep{Zhang2015HLTau,Okuzumi2016HLTau,Owen2020Snowlines} and MHD effects~\citep{Suriano2018,RiolsLesur2020,CuiBai2021}. Some of the mechanisms requires active participation from dust itself due to its back-reaction~\citep{Takahashi2014SGI,Takahashi2016SGI,Tominaga2019SGI,Tominaga2020SGI}. In all these scenarios, it is crucial to co-evolve gas and dust in a self-consistent manner to help constrain the physical mechanisms behind the observations.

Computationally, dust is commonly treated either as Lagrangian (super-) particles, or as pressureless fluids. The particle methods have been implemented in several MHD codes including Pencil~\citep{JohansenOishi2007}, Athena~\citep{Bai2010particle}, FARGO-ADSG~\citep{BaruteauZhu2016} and PLUTO~\citep{MignoneFlock2019Dust}. It has also been naturally employed in smoothed particle hydrodynamic (SPH) codes including PHANTOM~\citep{PricePHANTOM2018}. One major advantage of the Lagrangian treatment is being able to properly handle particle crossing, more relevant for particles that are marginally or loosely coupled to the gas, which is important for studying planetesimal formation by the SI. On the other hand, it is generally difficult to handle the highly stiff regime of extreme particle concentration (\citealp{Bai2010particle}, but see~\citealp{YangJohansen2016,Moseley2022}), and achieving good load balancing can be challenging for very large simulations~\cite[but see][]{JohansenKlahr2011}. Moreover, it is common to treat the unspecified source of disk turbulence as an effective viscosity in gas dynamic simulations. Doing so for particles can be involved, especially if one were to further consider dust back-reaction.

The alternative fluid treatment of dust is gaining popularity, such as in PIERNIK~\citep{Hanasz2010I,Hanasz2010II}, MPI-AMRVAC~\citep{Porth2014MPIAMRVAC,Xia2018MPIAMRVAC}, LA-COMPASS~\citep{LiLi2005Vorticity,LiLubow2009TypeI,FuLi2014} and FARGO3D~\citep{Benitez2016FARGO3D,Benitez2019FARGO3D}. This approach is more appropriate for relatively strongly coupled dust, as they quickly respond to fluid motion to minimize particle crossing. As separate fluids are co-located with gas in the computational domain, stiffness issues can be overcome by designing fully-implicit schemes for the drag source term simultaneously on gas and dust, and load balancing can be trivially satisfied. Dust diffusion can be easily handled by incorporating a concentration diffusion source term~\citep{Cuzzi1993,YoudinLithwick2007}. Finally, this approach is generalizable to further incorporate dust coagulation~\citep{LiLi2019DustGrowth,Drazkowska2019,LiLi2020Coagulation}, so that one can self-consistently compute the dust size distribution at every simulation cell.

In this paper, we describe the algorithm, implementation and numerical tests of a multifluid dust module in the Athena++ MHD code~\citep{Stone2020}. Our development features a set of dust integrators, particularly two fully-implicit integrators that can handle all stiff regimes while maintaining 2nd-order accuracy, which improve upon previous works which were either explicit, such as MPI-AMRVAC~\citep{Porth2014MPIAMRVAC}, FARGO-ADSG~\citep{BaruteauZhu2016} and PHANTOM~\citep{PricePHANTOM2018}, or implicit but only 1st-order accurate~\cite[FARGO3D,][]{Benitez2019FARGO3D}. With Athena++ being a Godunov MHD code, our implementation naturally conserves total momentum and energy. Moreover, we provide a consistent formulation of dust concentration diffusion, and show that additional correction terms in the momentum equations of dust are necessary to properly conserve total momentum and maintain Galilean invariance. Implementing these terms yield physically sensible results in a number of test problems.

The outline of this paper is as follows. In Section~\ref{sec:schemes}, we describe the equations, our numerical schemes and implementations. In Section~\ref{sec:tests}, we present the benchmark tests, including collisions between gas and dust, dust diffusion with or without momentum correction, linear and non-linear tests of the SI, global curvilinear simulations of the SI, as well as static/adaptive mesh refinement tests. Finally, we summarize and discuss our results in Section~\ref{sec:summary}.

\section{Numerical Scheme}~\label{sec:schemes}

In this section, we describe the basic equations including consistent formulation of dust concentration diffusion, as well as the numerical schemes and implementation of the multifluid dust module in Athena++.

\subsection{General Equations (Conservative Form)}~\label{subsec:equations}

We start by presenting the full set of equations of gas and multifluid dust. We use subscripts ``d'' and ``g'' to denote ``dust'' and ``gas''. Let there be $N_\tx{d}$ dust species, each characterized by a stopping time $T_{\tx{s},n}$ representing the timescale they respond to gas drag, where we use a label ``$n$'' for the $n$-th dust species. In conservative form, the equations read:

\begin{equation}
\frac{\pt \rho_{\tx{g}}}{\pt t} + \nabla \cdot \left(\rho_{\tx{g}} \mbs{v}_{\tx{g}}\right) =0\ ,
\label{eq:gas_con}
\end{equation}

\begin{equation}
\begin{aligned}
  \frac{\pt \left(\rho_{\tx{g}}\mbs{v}_{\tx{g}}\right)}{\pt t} &+  \nabla\cdot\left(\rho_{\tx{g}}\mbs{v}_{\tx{g}} \mbs{v}_{\tx{g}} + P_{\tx{g}}\mathsf{I}+\mbs{\Pi}_{\nu}\right)= \\
  & \rho_{\tx{g}} \mbs{f}_\tx{g,src}+\sum^{N_\tx{d}}_{n=1}\rho_{\tx{d},n} \frac{\mbs{v}_{\tx{d},n} - \mbs{v}_{\tx{g}}}{T_{\tx{s},n}}\ ,
\end{aligned}
\label{eq:gas_mom}
\end{equation}

\begin{equation}
\begin{aligned}
&\frac{\pt E_{\tx{g}}}{\pt t} + \nabla \cdot\left[\left(E_{\tx{g}}+P_{\tx{g}}\right) \mbs{v}_{\tx{g}}+\mbs{\Pi}_{\nu} \cdot \mbs{v}_{\tx{g}}\right]= \rho_{\tx{g}} \mbs{f}_\tx{g,src} \cdot \mbs{v}_{\tx{g}} \\
&+ \sum^{N_\tx{d}}_{n=1}\rho_{\tx{d},n} \frac{\mbs{v}_{\tx{d},n} - \mbs{v}_{\tx{g}}}{T_{\tx{s},n}}\cdot \mbs{v}_{\tx{g}} + \omega\sum^{N_\tx{d}}_{n=1}\rho_{\tx{d},n} \frac{\left(\mbs{v}_{\tx{d},n} - \mbs{v}_{\tx{g}}\right)^2}{T_{\tx{s},n}}\ ,
\end{aligned}
\label{eq:gas_erg}
\end{equation}

\begin{equation}
  \frac{\pt \rho_{\tx{d},n}}{\pt t} +\nabla \cdot \left(\rho_{\tx{d},n} \mbs{v}_{\tx{d},n} + \mbs{\mrs{F}}_{\tx{dif},n}\right) = 0\ ,
\label{eq:dust_con}
\end{equation}

\begin{equation}
\begin{aligned}
  \frac{\pt \rho_{\tx{d},n} \left(\mbs{v}_{\tx{d},n} + \mbs{v}_{\tx{d,dif},n}\right)}{\pt t} &+ \nabla\cdot(\rho_{\tx{d},n}\mbs{v}_{\tx{d},n} \mbs{v}_{\tx{d},n} + \mbs{\Pi}_{\tx{dif},n}) = \\
   \rho_{\tx{d},n} &\mbs{f}_{\tx{d,src},n} + \rho_{\tx{d},n}\frac{\mbs{v}_{\tx{g}} - \mbs{v}_{\tx{d},n}}{T_{\tx{s},n}}\ .
\end{aligned}
\label{eq:dust_mom}
\end{equation}

There are $4N_\tx{d} + 5$ equations in total, where Equations (\ref{eq:gas_con}) to (\ref{eq:gas_erg}) are the gas continuity, momentum and energy equations, and Equations (\ref{eq:dust_con}), (\ref{eq:dust_mom}) are the continuity and momentum equations for the dust species, which are treated as pressureless fluids~\citep{GaraudBarriere2004DustFluids}. In the above, $\rho$ is the density, $\mbs{v}$ is the velocity, $P_\tx{g}$ is the gas pressure, $\mathsf{I}$ is the identity tensor, and $E_\tx{g}=(\rho_\tx{g} \mbs{v}_\tx{g}^2)/2+P_\tx{g}/(\gamma-1)$ is the total energy density of gas, with $\gamma$ being the adiabatic index. Here we neglect magnetic fields, thermal conduction, etc., as they do not directly couple to dust, and our dust fluid module is fully compatible with these existing features.\footnote{We treat dust fluids as neutral, but future extensions may incorporate dust charge, e.g.,~\citealp{HopkinsSquire18}.}

We incorporate gas viscosity which mimics the presence of external turbulence, described by the viscous stress tensor $\mbs{\Pi}_{\nu}$:
\begin{equation}
  \Pi_{\nu,ij}=\rho_{\tx{g}} \nu_\tx{g}\left(\frac{\pt v_{\tx{g},i}}{\pt x_{\tx{g},j}}+\frac{\pt v_{\tx{g},j}}{\pt x_{\tx{g},i}}-\frac{2}{3} \delta_{ij} \nabla \cdot \mbs{v_{\tx{g}}}\right)\ ,
\end{equation}
where $\nu_\tx{g}$ is the kinematic viscosity. Closely related to the gas viscosity is a dust diffusivity $D_{\tx{d},n}$, which leads to concentration diffusion. We treat the diffusivity for each dust species as a free parameter to be specified by the user, while they are usually prescribed as~\citep{YoudinLithwick2007}:

\begin{equation}
  D_{\tx{d},n} = \frac{\nu_{\tx{g}}}{1+\left(T_{\tx{s},n}/T_{\tx{g,eddy}}\right)^2}\ ,
\end{equation}
where $T_{\rm g,eddy}$ is the turbulent eddy time of the external turbulence. With this, the dust concentration diffusion flux, acting on the dust continuity equation, is given by
\begin{equation}
  \mbs{\mrs{F}}_{\tx{dif},n} \equiv -\rho_{\tx{g}} D_{\tx{d},n} \nabla \left(\frac{\rho_{\tx{d},n}}{\rho_{\tx{g}}}\right) = \rho_{\tx{d},n}\mbs{v}_{\tx{d,dif},n}\ ,\label{eq:conflx}
\end{equation}
which also gives the definition of the effective dust drift speed $\mbs{v}_{\tx{d,dif},n}$ due to concentration diffusion. Associated with this concentration diffusion, correction terms must be incorporated to the dust momentum equation to ensure consistent momentum diffusion flux~\citep{Tominaga2019SGI} and Galilean invariance. The individual components of the momentum diffusion flux tensor are given by
\begin{equation}
  \Pi_{\tx{dif},n,ij}=v_{\tx{d},n,j} \mrs{F}_{\tx{dif},n,i} + v_{\tx{d},n,i} \mrs{F}_{\tx{dif},n,j}\ .\label{eq:momflx}
\end{equation}
Full derivations of the concentration diffusion terms will be presented in Section~\ref{ssec:mom_correction}.

The last terms of the right hand sides of the momentum equations (\ref{eq:gas_mom}) and (\ref{eq:dust_mom}) correspond to the aerodynamic drag between gas and dust. 
Here we assume linear drag law, where $T_{{\rm s},n}$ is independent of velocity.
Additional two source terms are added to the energy equation (\ref{eq:gas_erg}), which correspond to the work done by the drag, and frictional heating. We have included a parameter $\omega$ to control the level of frictional heating, being $0$ to be turned off, and $1$ when all dissipation is deposited to the gas.\footnote{In reality, some of the dissipation must lead to heating of the dust. If assuming gas and dust should maintain the same temperature, one should assign $\omega=c_{V,\tx{g}}\rho_\tx{g}/(c_{V,\tx{g}}\rho_\tx{g}+c_{V,\tx{d}}\rho_\tx{d})$, where $c_{V,\tx{g}}$ and $c_{V,\tx{d}}$ are the heat capacity of gas and dust, respectively, and $\rho_\tx{d}=\sum_{n=1}^{N_\tx{d}}\rho_{\tx{d},n}$.}

Other external source terms are denoted by $\mbs{f}_\tx{src}$, which may include stellar and/or planetary gravity in disk problems depending on applications. They are implemented as explicit source terms added on the momentum equations (\ref{eq:gas_mom}) and (\ref{eq:dust_mom}) following the standard in Athena++~\citep{Stone2020}. Associated with them is a source term $W \equiv \mbs{f}_\tx{g,src} \cdot \mbs{v}_\tx{g}$ in the energy equation accounting for the work done by the source terms.

Note that our formulation does not contain an energy equation for dust, thus does not ensure global energy conservation of the composite dust-gas system. While this is not of overwhelming concern in typical applications, future generalization to incorporate a dust energy equation is possible. At algorithmic level, we thus aim at full momentum conservation, and implement energy source terms to match the overall accuracy of the algorithm.

\subsection{Consistent Formulation of Dust Concentration Diffusion}~\label{ssec:mom_correction}

Here we derive dust fluid equations in the presence of turbulent diffusion, following the procedures of~\cite{Cuzzi1993} and~\cite{Tominaga2019SGI}. We use the Reynolds averaging technique with approximate closure relations to properly account for the role of turbulence at sub-grid level while preserving global conservation laws. In doing so, we are interested in the physics on time (and potentially length) scales above those for the turbulence, and hence any physical variable $A$ is decomposed into a time-averaged part $\ovl{A}$ and a fluctuating part $\Delta A$, i.e., $A = \ovl{A} + \Delta A$.

Without loss of generality, we focus on a single dust species and drop its label $n$. We start from the standard dust fluid equations in conservation form:
\begin{equation}
\begin{aligned}
\frac{\pt \rho_{\tx{d}}}{\pt t}+\nabla \cdot\left(\rho_{\tx{d}} \mbs{v}_{\tx{d}}\right) &=0\ , \\
\frac{\pt \rho_{\tx{d}} \mbs{v}_{\tx{d}}}{\pt t}+\nabla \cdot\left(\rho_{\tx{d}} \mbs{v}_{\tx{d}}  \mbs{v}_{\tx{d}}\right) &=\rho_\tx{d} \frac{\mbs{v}_\tx{g}-\mbs{v}_\tx{d}}{T_\tx{s}}\ .
\end{aligned}
\end{equation}
Taking averages to the continuity equation, we obtain
\begin{equation}
\begin{aligned}
  \frac{\pt \ovl{\rho_{\tx{d}}}}{\pt t}+\nabla \cdot\left(\ovl{\rho}_{\tx{d}} \ovl{\mbs{v}_{\tx{d}}}+\ovl{\Delta \rho_{\tx{d}} \Delta \mbs{v}_{\tx{d}}}\right)=0\ .
\end{aligned}
\end{equation}
The extra term $\ovl{\Delta \rho_{\tx{d}} \Delta \mbs{v}_{\tx{d}}}$, by definition, corresponds to dust concentration diffusion flux (\ref{eq:conflx})
\begin{equation}
  \ovl{\Delta \rho_{\tx{d}} \Delta \mbs{v}_{\tx{d}}} \equiv \mbs{\mrs{F}}_{\tx{dif}}
  = \ovl{\rho_{\tx{d}}}\mbs{v}_{\tx{d},\tx{dif}}\ .
\end{equation}

Next, by taking averages to the momentum equation, we obtain
\begin{equation}
\begin{aligned}
  &\frac{\pt \ovl{\rho}_{\tx{d}} \ovl{\mbs{v}_{\tx{d},j}}}{\pt t}+ \frac{\pt \ovl{\Delta \rho_{\tx{d}} \Delta \mbs{v}_{\tx{d},j}}}{\pt t}+ \frac{\pt}{\pt x_{i}}(\ovl{\rho_{\tx{d}}} \ovl{\mbs{v}_{\tx{d},i}} \ovl{\mbs{v}_{\tx{d},j}}\\
  &+\ovl{\Delta \rho_{\tx{d}} \Delta \mbs{v}_{\tx{d},i}} \ovl{\mbs{v}_{\tx{d},j}}+\ovl{\mbs{v}_{\tx{d},i}} \ovl{\Delta \rho_{\tx{d}} \Delta \mbs{v}_{\tx{d},j}}+\ovl{\rho_{\tx{d}}} \ovl{\Delta \mbs{v}_{\tx{d},i} \Delta \mbs{v}_{\tx{d},j}}) \\
  &=\ovl{\rho_{\tx{d}}} \frac{\ovl{\mbs{v}_{\tx{g},j}}-\ovl{\mbs{v}_{\tx{d},j}}}{T_\tx{s}}+\frac{\ovl{\Delta \rho_{\tx{d}} \Delta \mbs{v}_{\tx{g},j}}-\ovl{\Delta \rho_{\tx{d}} \Delta \mbs{v}_{\tx{d},j}}}{T_\tx{s}}\ .
\end{aligned}
\end{equation}
At this stage, it is often argued that one can drop the second term on the left assuming the time-dependent diffusion flux is small compared to that of the bulk flow~\citep{Cuzzi1993,Tominaga2019SGI}.
However, our analysis shows that this would violate Galilean invariance (see Appendix~\ref{app:galilean}, and also numerical tests in Section~\ref{subsec:diffusion}), and hence it must be kept. The second and third terms in the momentum flux can be reduced using the effective dust drift velocity $\mbs{v}_{\tx{d,dif}}$, which leads to the expression of momentum diffusion flux (\ref{eq:momflx}). We note that momentum conservation does not necessarily requires the inclusion of momentum diffusion flux, but this flux is important when considering angular momentum conservation in disk problems \citep{Tominaga2019SGI}, as well as for ensuring Galilean invariance.

For the last term in the momentum flux $\ovl{\rho}_{\tx{d}} \ovl{\Delta \mbs{v}_{\tx{d},i} \Delta \mbs{v}_{\tx{d},j}}$, we may use the simple closure relation by~\cite{ShariffCuzzi2011} and~\cite{Tominaga2019SGI} as
\begin{equation}
  \ovl{\Delta \mbs{v}_{\tx{d},i} \Delta \mbs{v}_{\tx{d},j}} = \delta_{ij}c_\tx{s,d}^2\ .
\end{equation}
where $c_\tx{s,d}$ is the effective dust sound speed. This term can be neglected in the multifluid approach~\citep{GaraudBarriere2004DustFluids}. The second term on the right hand side is also neglected with the expectation that the standard drag term dominates, as in~\cite{Tominaga2019SGI}.

With all these considerations, we recover the dust momentum equation shown in Section~\ref{subsec:equations}, here rewritten as
\begin{equation}
\begin{aligned}\label{eq:dust_mom1}
  \frac{\pt \rho_{\tx{d}}\left(v_{\tx{d},j}+v_{\tx{d,dif},j}\right)}{\pt t}&+ \frac{\pt}{\pt x_{i}}(\rho_{\tx{d}} v_{\tx{d},i}v_{\tx{d},j}+\rho_{\tx{d}}v_{\tx{d,dif},i}v_{\tx{d},j}\\
  +\rho_{\tx{d}}v_{\tx{d},i}v_{\tx{d,dif},j})&=\rho_{\tx{d}} \frac{v_{\tx{g},j}-v_{\tx{d},j}}{T_\tx{s}}\ .
\end{aligned}
\end{equation}
where for notational convenience, we can drop the overline and interpret the dust fluid quantities in the averaged sense. The presence of time-derivative on $\rho_d\mbs{v}_\tx{d,dif}$ in the momentum equation is the inevitable consequence of this averaging procedure. Missing this term would lead to unphysical behaviors as we demonstrate in Section \ref{subsec:momentum_correction}. Implementing this term also requires special care, as will be discussed in Section~\ref{subsec:diffusion}. 

\subsection{Dust-Gas Drag Integrators}~\label{subsec:drags}

The drag term involves interactions between gas and all dust species. As a special source term to both gas and dust, the drag integrator aims to solve the following equation
\begin{equation}
\centering
\begin{aligned}
\frac{\pt \mbs{M}}{\pt t} =
\begin{bmatrix}
\sum^{N_\tx{d}}_{n=1} \alpha_n\left( \mbs{M}_{\tx{d},n}- \epsilon_n \mbs{M}_\tx{g} \right)\\
\alpha_1 (\epsilon_1 \mbs{M}_\tx{g} - \mbs{M}_{\tx{d},1})\\
\alpha_2(\epsilon_2 \mbs{M}_\tx{g} - \mbs{M}_{\tx{d},2})\\
\vdots \\
\alpha_n(\epsilon_n \mbs{M}_\tx{g} - \mbs{M}_{\tx{d},n})\\
\end{bmatrix}
\equiv \mbs{f}_\tx{drag} \left(\mbs{M}, \mbs{W}\right)\ ,
\end{aligned}
\label{eq:drag_force}
\end{equation}
with $\mbs{f}_\tx{drag}$ being the mutual drag force,  $\mbs{M} \equiv [\mbs{M}_\tx{g},$ $\mbs{M}_\tx{d,1},\ \dots,\ \mbs{M}_{\tx{d},n}]^\top = [\rho_\tx{g} \mbs{v}_\tx{g},\ \rho_{\tx{d},1} \mbs{v}_\tx{d,1},\ \dots,\ \rho_{\tx{d},n} \mbs{v}_{\tx{d},n}]^\top$ is the momentum vector of gas and dust. The remaining variables are denoted as $\mbs{W}$ given by ($\mbs{\epsilon} , \mbs{\alpha})$, where $\mbs{\epsilon} = \left[\epsilon_1, \dots, \epsilon_n \right] \equiv [\rho_{\tx{d},\ 1}/\rho_\tx{g},\ \dots,\ \rho_{\tx{d},n}/\rho_\tx{g} ]$, and $\mbs{\alpha} = \left[\alpha_1, \dots \alpha_n\right] \equiv [T_{\tx{s},1}^{-1}, \dots T_{\tx{s},n}^{-1}]$. They are treated as constant parameters in the integrator.

The drag term is potentially stiff in two regimes. First, when the dust stopping time $T_s$ is very small, and stiffness arises when $T_S<\Delta t\equiv h$, the hydrodynamic time step. Second, when $\sum_n\epsilon_n\gg1$, which arises when dust is strongly concentrated. The stiff regimes should be handled by fully-implicit integrators for stability. We note that for particle-based methods, handling the first regime of stiffness is relatively straightforward~\citep{Bai2010particle,FungMuley2019,MignoneFlock2019Dust}, whereas handling the second regime requires extra care, where one either artificially reduces particle back-reaction~\citep{Bai2010particle}, or sacrifice the time step~\citep{LiYoudin2021}, and more rigorous treatment demands substantially more computational cost~\citep{YangJohansen2016}. With the fluid-treatment of dust, one can directly solve the above equation implicitly, which automatically handles both stiffness regimes~\citep{Benitez2019FARGO3D}.
In doing so, we need to evaluate the Jacobian of $\mbs{f}_\tx{drag}$, and for brevity we drop the subscript ``drag'':
\begin{equation}
\centering
\label{eq:jacobian}
\begin{aligned}
  \frac{\pt \mbs{f}}{\pt \mbs{M}}=
\begin{bmatrix}
-\sum^{N_\tx{d}}_n \epsilon_i \alpha_i & \alpha_1  & \alpha_2  & \cdots & \alpha_n  \\
\epsilon_1 \alpha_1           & -\alpha_1 & 0         & \cdots & 0         \\
\epsilon_2 \alpha_2           & 0         & -\alpha_2 & \cdots & 0         \\
\vdots                        & \vdots    & \vdots    & \ddots & \vdots    \\
\epsilon_n \alpha_n           & 0         & 0         & \cdots & -\alpha_n \\
\end{bmatrix}\ .
\end{aligned}
\end{equation}
Note that for linear drag law, this Jacobian applies to individual dimensions (which are independent of each other).

We have implemented a number of different drag integrators.
Here we describe the algorithms of fully-implicit integrators that we develop for achieving numerical stability and towards higher-order accuracy. The implementation of other simpler integrators, which are explicit or semi-implicit that are useful for non-stiff problems, are described in Appendix~\ref{app:otherdrag}.

\subsubsection{First Order Fully-Implicit Method}~\label{subsec:rk1im}

We start from the standard backward Euler method, which is a single-stage integrator to be combined with the Runge-Kutta 1 (RK1, or forward Euler) time integrator in Athena++. Integrating from step $n$ to $n+1$, the format is given by

\begin{equation}
  \begin{aligned}
    \mbs{M}^{(n+1)}&= \mbs{M}^{(n)} + h \mbs{f}\left(\mbs{M}^{(n+1)}, \mbs{W}^{(n)}\right)\ .\\
  \end{aligned}
  \label{app-eq:rk1im}
\end{equation}
Note that this numerical format guarantees momentum conservation.
Substituting
\begin{equation}
\begin{aligned}
\mbs{f}(\mbs{M}^{(n+1)}, \mbs{W}^{(n)})&= \mbs{f}(\mbs{M}^{(n)}, \mbs{W}^{(n)})\\
&+ \frac{\pt \mbs{f}}{\pt \mbs{M}}|^{(n)}(\mbs{M}^{(n+1)} - \mbs{M}^{(n)})\ ,\\
\end{aligned}
\end{equation}
we can update the momentum $\Delta\mbs{M}\equiv\mbs{M}^{(n+1)} - \mbs{M}^{(n)}$ by
\begin{equation}
    \Delta\mbs{M}=
    \left(\mathsf{I} - h \frac{\pt \mbs{f}}{\pt \mbs{M}}|^{(n)}\right)^{-1} h \mbs{f} \left(\mbs{M}^{(n)}, \mbs{W}^{(n)}\right)\ ,\\
\end{equation}
where $\mathsf{I}$ is the identity matrix and evaluating $\Delta\mbs{M}$ involves matrix inversion.
This is the main integrator implemented in FARGO3D~\citep{Benitez2019FARGO3D}, which makes the mutual drag interaction unconditional stable, despite of being only 1st-order accurate in time.
With the simple form of the Jacobian (\ref{eq:jacobian}), the matrix in the backward Euler method can be solved efficiently on order $\sim O(N_d)$ instead of $\sim O(N_d^3)$ as in standard LU decomposition \citep{KrappBenitez2020RNAAS}.

The energy source term on the gas has two parts. The first arises from the work done by the drag force. To better preserve energy conservation, this term should be implemented as the change in the gas kinetic energy due to gas drag:

\begin{equation}\label{eq:eng1}
  \Delta E_{\tx{g},1}=\Delta \mbs{M}_\tx{g} \cdot (\mbs{v}^{(n)}_\tx{g}+\mbs{v}^{(n+1)}_\tx{g})/2\ ,
\end{equation}
The second part is from frictional heating~\citep{Marble1970,LaibePrice2014One,MignoneFlock2019Dust}, which is associated with the reduction of total kinetic energy in the gas-dust system. This can be calculated by
\begin{equation}\label{eq:eng2}
  \Delta E_{\tx{g},2}= \Delta \mbs{M}_\tx{g} \cdot \frac{\mbs{v}^{(n)}_\tx{g}+\mbs{v}^{(n+1)}_\tx{g}}{2}
  +\sum_{n=1}^{N_\tx{d}} \Delta \mbs{M}_{\tx{d},n} \cdot \frac{\mbs{v}^{(n)}_{\tx{d},n}+\mbs{v}^{(n+1)}_{\tx{d},n}}{2}\ ,
\end{equation}
The source terms for the energy equation should thus be
\begin{equation}\label{eq:eng3}
  E_\tx{g}^{(n+1)}=E_\tx{g}^{(n)}
    +\Delta E_{\tx{g},1}-\omega\Delta E_{\tx{g},2}\ .
\end{equation}

\subsubsection{Second Order Fully-Implicit Methods}\label{subsubsec:2ndIm}

Next we build two fully-implicit drag integrators to be combined with the van-Leer 2 (VL2) and the Runge-Kutta 2 (RK2) time integrators in Athena++. We refer to them as the ``VL2-Implicit'' and ``RK2-Implicit'' integrators, respectively. Both integrators involve two stages.
Here we describe their implementation, while the derivation of the algorithm can be found in Appendix~\ref{app:imp}.

\subsubsection*{VL2-Implicit}

Stage I: We apply the backward Euler method to update the system momenta from step $n$ for half a time step $h/2$, denoted by a prime $'$:

\begin{equation}
\label{eq:VL2IMPLICIT-1}
    \Delta\mbs{M}'=
    \left(\mathsf{I} -\frac{h}{2}\frac{\pt \mbs{f}}{\pt \mbs{M}}|^{(n)}\right)^{-1}\frac{h}{2}\mbs{f} \left(\mbs{M}^{(n)}, \mbs{W}^{(n)}\right)\ ,\\
\end{equation}
Matrix inversion in this stage can be similarly achieved on order $O(N_d)$.
The update in gas energy at this stage is exactly analogous to that in the backward Euler method, which we do not repeat.

Stage II: the momentum is updated from step $n$ to $n+1$ using the following
\begin{equation}
\label{eq:VL2IMPLICIT-2}
\begin{aligned}
  \Delta \mbs{M}&=\mbs{\Lambda}^{-1} \left(\mathsf{I} - \frac{h}{2} \frac{\pt \mbs{f}}{\pt \mbs{M}}\bigg|'\right) h \mbs{f}\left(\mbs{M}^{(n)}, \mbs{W}'\right)\ ,\\
\end{aligned}
\end{equation}
where
\begin{equation}\label{eq:LamVL2}
  \mbs{\Lambda} \equiv \mathsf{I} - \left(\mathsf{I} - \frac{h}{2} \frac{\pt \mbs{f}}{\pt \mbs{M}}\bigg|'\right) h \frac{\pt \mbs{f}}{\pt \mbs{M}}|^{(n) }\ .
\end{equation}
Note that this matrix is more complex and should be inverted by LU decomposition~\citep{Press1986numerical}.
The update in gas energy has exactly the same form as Equations (\ref{eq:eng1}) to (\ref{eq:eng3}), which we do not repeat.

\subsubsection*{RK2-Implicit}

Stage I: We use the backward Euler method with time step $h$ to calculate the momentum at step $n+1$, which is exactly the same as described in Section~\ref{subsec:rk1im}. We still denote the quantities at the end of this stage using a prime $'$.

Stage II: The momentum at stage $n+1$ is:
\begin{equation}
\label{eq:RK2IMPLICIT-2}
\begin{aligned}
  \Delta \mbs{M}=
 \mbs{\Lambda}^{-1} &\bigg[h \mbs{f}(\mbs{M}^{(n)}, \mbs{W}') \\
 &+\bigg(\mathsf{I} - h \frac{\pt \mbs{f}}{\pt \mbs{M}}\bigg|'\bigg) h \mbs{f}(\mbs{M}^{(n)}, \mbs{W}^{(n)})\bigg]\ ,
\end{aligned}
\end{equation}
where
\begin{equation}
\begin{aligned}\label{eq:LamRK2}
\mbs{\Lambda} &\equiv \mathsf{I} - h \frac{\pt \mbs{f}}{\pt \mbs{M}}\bigg|^{(n)} + \frac{h^2}{2} \frac{\pt \mbs{f}}{\pt \mbs{M}}\bigg|' \frac{\pt \mbs{f}}{\pt \mbs{M}}\bigg|^{(n)}\ .
\end{aligned}
\end{equation}

Similarly, matrix inversion is solved by LU decomposition.
The update in gas energy also has exactly the same form as Equations (\ref{eq:eng1}) to (\ref{eq:eng3}), which we do not repeat.

Note that this integration scheme is in essence the same as the fully-implicit particle integrator in~\cite{Bai2010particle}.

\subsubsection[]{Coupling Explicit Hydrodynamic Integrators with Implicit Drag Integrators}\label{sssec:combine}

Special care must be taken when combining implicit integrators with the explicit hydrodynamic integrators and source terms. When they are treated separately, the combined algorithm would only be 1st-order accurate,\footnote{The implicit-explicit Runge-Kutta schemes are viable choices~\citep{Pareschi05}, they usually involve more stages of integration than the order of accuracy achieved, and do not necessarily match the existing hydrodynamic integrators in Athena++.} and the implicit drag integrator cannot maintain exact equilibrium solutions.

To overcome these issues, we may consider the advection, diffusion and other hydrodynamic source terms as an ``add-on'' to $\mbs{f}$. In other words, $\mbs{f}$ in the drag algorithms above represents the combination of the drag force ($\mbs{f}_{\rm drag}$, treated implicitly), as well as other explicit terms including advection and other source terms $\mbs{G}_M$ acting on the gas and dust momenta
\begin{equation}
    \mbs{f}\equiv\mbs{f}_{\rm drag}(\mbs{M},\mbs{W})+\mbs{G}_M(\mbs{\overline{U}})\ ,
\end{equation}
where $\mbs{G}_M$ is expressed in terms of conserved variables $\mbs{U}$. By adding an overline on $\mbs{U}$, we treat these other explicit terms as known constant, readily obtained in the hydrodynamic integrator. In the hydrodynamic integration from step 1 to step 2 over time interval $\Delta t$, we estimate $\mbs{G}_M$ to be the momentum update from explicit terms
\begin{equation}
    \mbs{G}_M(\mbs{\overline{U}})=\frac{\mbs{M}^{(2)}-\mbs{M}^{(1)}}{\Delta t}\ ,
\end{equation}
where $\mbs{M}^{(2),(1)}$ represents the momenta before and after explicit integration steps (advection, diffusion and other explicit source terms). By treating this term as a constant, the Jacobian and the $\Lambda$ matrices described in the previous subsection remain unchanged.

Implementing the above requires extra storage to store the momentum updates, and that we must finish all explicit steps in the hydrodynamic integration before entering the drag integrator. We will show that our approach successfully achieves 2nd-order accuracy when using VL2-Implicit and RK2-Implicit integrators, and it also allows us to achieve exact equilibrium solutions involving the drag force.

\subsection[]{Integration of Multifluid Dust Equations}

The integration of dust fluid is divided into several parts (advection, diffusion, source terms and drag). Except for the drag term (described in the previous subsection), the other terms are treated independently and explicitly, and we describe their implementation in this subsection.

\subsubsection{General Procedures}~\label{subsubsec:procedures}

Following the standard routine in Athena++, each integration time step is divided into a number of stages depending on the time integrator employed (see detailed descriptions in~\citealp{Stone2020}). In each stage, the integration procedures involve updating conserved variables based on primitive variables by evolving the fluid equations by $dt$. Our multifluid dust module supports Athena++ time integrators up to second order, including 1st-order Runge-Kutta (RK1), 2nd-order Runge-Kutta (RK2) and van-Leer integrator (VL2).

For each dust species, the primitive ($\mbf{W}_\tx{d}$) and conserved ($\mbf{U}_\tx{d}$) variables are
\begin{equation}
\mbf{W}_{\tx{d}}=
\begin{bmatrix}
\rho_\tx{d}\\
\mbs{v}_\tx{d}\\
\end{bmatrix}
\ ,\quad
\mbf{U}_{\tx{d}}=
\begin{bmatrix}
\rho_\tx{d}\\
\rho_\tx{d}(\mbs{v}_\tx{d}+\mbs{v}_\tx{d,dif})\\
\end{bmatrix}\ .
\label{eq:dustvar}
\end{equation}
Note that the presence of time derivative on $\rho_\tx{d}\mbs{v}_\tx{d,dif}$ in the momentum equation suggests that the concentration diffusion flux should be considered as part of the conserved dust momentum. The total momentum is thus $\rho\mbs{v}_\tx{g}+\sum_i\rho_{\tx{d},i}(\mbs{v}_{\tx{d},i}+\mbs{v}_{\tx{d,dif},i})$.

Integrating the bulk part of the dust fluid is very similar that of hydrodynamics in Athena++. The main procedures involves the reconstruction of primitive variables at cell interfaces, followed by solving a Riemann problem to obtain the mass and momentum fluxes, after which we update the dust fluid quantities from flux gradients. Same as in Athena++, the multifluid dust module supports spatial reconstructions up to third order.

As pressureless fluids, the Riemann problem for dust fluids is greatly simplified. In one-dimension along the $x$-direction, given the left/right states $\mbf{W}_{\tx{d}}^{L/R}$, we provide the Riemann flux for conserved variables as follows. The density flux reads
\begin{equation}
  F_x(\rho_d) =
  \begin{cases}
  \rho_{\tx{d}}^{L}v_{\tx{d},x}^{L} & v_{\tx{d},x}^{L}>0\ ,\ v_{\tx{d},x}^{R}>0\ , \\
  \rho_{\tx{d}}^{R}v_{\tx{d},x}^{R} & v_{\tx{d},x}^{L}<0\ ,\ v_{\tx{d},x}^{R}<0\ , \\
  0 & v_{\tx{d},x}^{L}<0\ ,\ v_{\tx{d},x}^{R}>0\ , \\
  \rho_{\tx{d}}^{L}v_{\tx{d},x}^{L}+\rho_{\tx{d}}^{R}v_{\tx{d},x}^{R} & v_{\tx{d},x}^{L}>0\ ,\ v_{\tx{d},x}^{R}<0\ . \\
  \end{cases}
\end{equation}

Similar expressions hold for the momentum flux for all three directions. Essentially, we use the upwind flux when the normal velocity in the L/R states are the same, set the flux to be zero when the L/R normal velocities diverge, and sum up the fluxes from the two sides when L/R normal velocities converge. The last treatment reflects that as pressureless fluids, the flows on the two sides can penetrate each other, just as particles.\footnote{Note that penetration is still prohibited within each cell, where dust fluid velocities get well mixed. Alternatively, one may set the flux to zero in this case. We do not find much practical differences in test problems by using different Riemann solvers for dust.}

The implementation of other source terms on dust, such as stellar gravity and source terms in shearing-box, as well as geometric source terms in cylindrical and spherical coordinates, are same as that of gas, which are treated explicitly.

\subsubsection{Dust Diffusion}~\label{subsec:diffusion}

The implementation of dust concentration diffusion starts by computing the concentration diffusion flux according to Equation (\ref{eq:conflx}). The fluxes are computed by standard finite differencing, and are located at cell interfaces. Next, we calculate the momentum diffusion flux according to Equation (\ref{eq:momflx}). This term contains two parts. The first part, $v_{\tx{d},n,j} \mrs{F}_{\tx{dif},n,i}$, describes the diffusion of the $j$-momentum in the $i$-direction. At the implementation level, its value is obtained by averaging from the upwind side based on the sign of the concentration diffusion flux $\mrs{F}_{\tx{dif},n,i}$. The second part, $v_{\tx{d},n,i} \mrs{F}_{\tx{dif},n,j}$, represents the advection of the $j$-diffusion flux in the $i$-direction. Its value is obtained by averaging from the upwind side based on the sign of the advection velocity $v_{\tx{d},n,i}$. In addition, we note that in cylindrical/spherical coordinates, we need to add extra diffusive geometric sources terms on the momentum and energy equations~\citep{SkinnerOstriker2010}.

Finally, we compute the concentration diffusion momenta and compare to the original concentration diffusion flux, from which we can estimate the contribution from the $\partial(\rho_{\tx{d},n}\mbs{v}_{\tx{d,dif},n})/\partial t$ term. The concentration diffusion momenta are stored in the cell center and are averaged by the nearby face-centered concentration diffusion fluxes. We note that although our formulation is Galilean invariant, it is not invariant to machine precision at implementation level, but the incorporation of this term is important to ensure approximate Galilean invariance in simulations.

\subsection[]{Flow chart}~\label{subsec:flowchart}

\begin{figure*}[htp]
\centering
\includegraphics[scale=0.18]{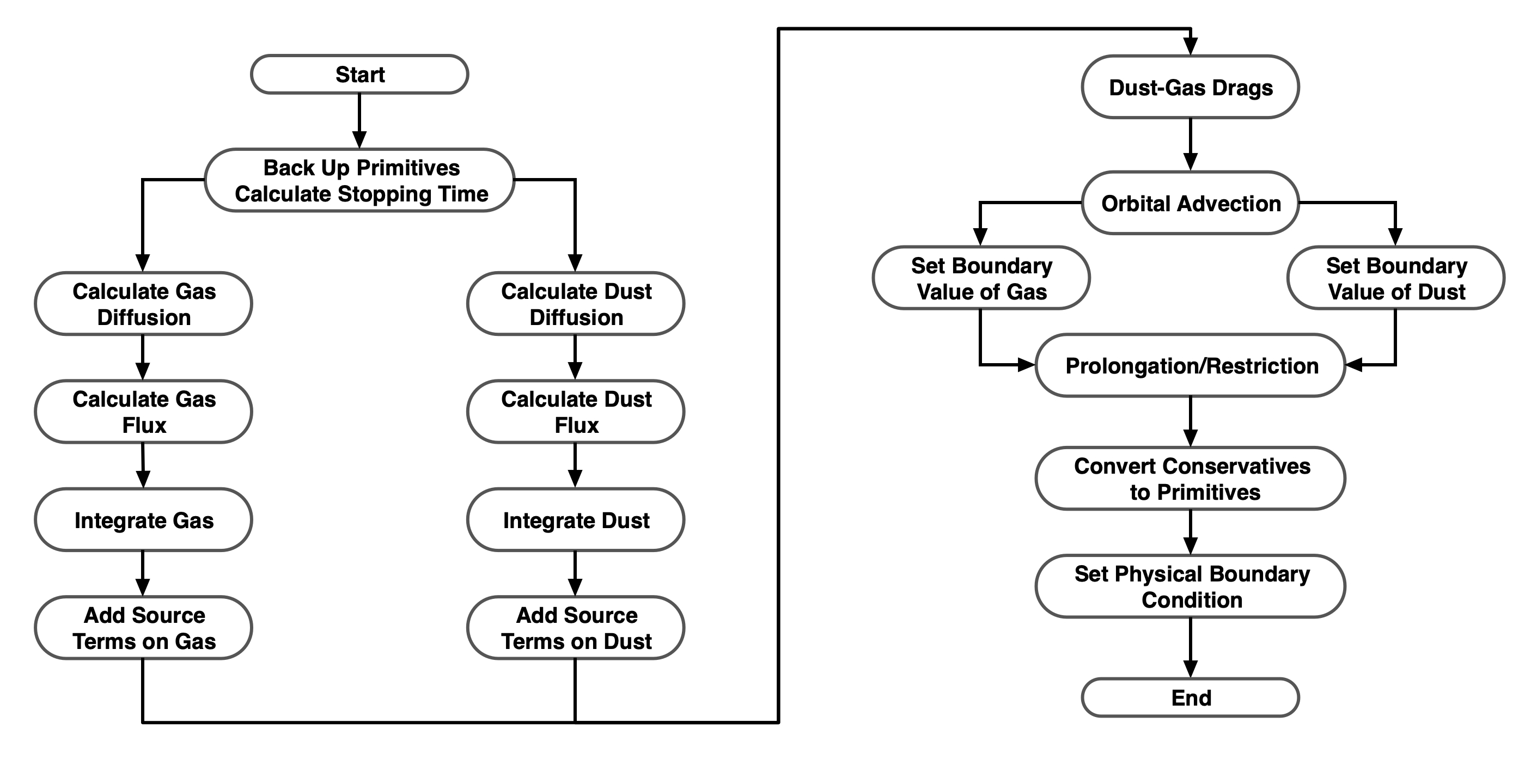}
\caption{Flow chart of a single integration stage of the multifluid dust module in Athena++.}
\label{fig:flowchart}
\end{figure*}

Figure~\ref{fig:flowchart} shows the flow chart of our multifluid dust module in Athena++, and we summarize the main steps over one integration stage below.

Step 1: Backup the primitive variables for both gas and dust and calculate the dust stopping time. The backed up primitive variables are used in the semi-implicit and fully-implicit drag integrators to ensure higher-order accuracy of the combined algorithm, as discussed in Section~\ref{sssec:combine}.

Step 2: Calculate the diffusion processes of gas and dust when applicable, including viscosity, thermal conduction, and resistivity on the gas, and concentration diffusion and momentum correction on dust fluids.

Step 3: Calculate the Riemann fluxes of both gas and dust, and integrate gas and dust fluids by applying flux divergence. Send and receive flux corrections when necessary for mesh refinement.

Step 4: Add explicit source terms on gas and dust, including geometric source terms for curvilinear coordinates.

Step 5: Apply any of the drag integrators, and use the backed-up variables to enhance the accuracy in implicit schemes.

Step 6: Do orbital advection when necessary (for disk problems).

Step 7: Send and receive boundary data, set boundary conditions, and do prolongation/restriction for mesh refinement.

Step 8: Convert conserved variables to primitive variables. When the dust momentum correction is turned on, the concentration diffusion flux calculated by Step 2 will be subtracted from the dust momenta.

After finishing all stages of an integration cycle, we calculate the new time step based on the Courant–Friedrichs–Lewy (CFL) condition for both gas and dust. The dust CFL condition is set according to maximum dust velocity and dust diffusion coefficient $D_\tx{d}$ in the same way as gas velocity and viscosity.

As a dust fluid module, it has fixed amount of floating point operations per meshblock per integration cycle, as opposed to particle-based approaches. Taking the advantage of the task-based execution model with excellent scalability of Athena++, our dust fluid module primarily adds a fixed fraction of computational cost. Such cost increases with $N_d$ non-linearly when using higher-order fully implicit drag solvers due to the matrix inversion whose cost scales as $O[(N_d+1)^3]$. In practice, we find that linear scaling approximately applies for $N_d\lesssim5$ and the cost of the drag solver is no more than the cost from rest of the dust integration scheme for $N_d\lesssim10$. Further details about code performance are provided in Appendix~\ref{app:performance}.

\section{Code Tests}~\label{sec:tests}

In this section, we show benchmark numerical tests of our multifluid dust module. They include the collisions between gas and dust, dust diffusion with momentum correction, linear/non-linear streaming instability and (static/adaptive) mesh refinement. We also follow the same dusty sound wave and dusty shock tests in Section 3.2 and 3.3 of~\cite{Benitez2019FARGO3D}.
To avoid repetitions, we show the test results of dusty sound wave and dusty shock in Appendix~\ref{app:dustysoundwave} and~\ref{app:dustyshock}. They demonstrate that our multi-fluid dust code achieves full second-order accuracy when coupled with hydrodynamics, and it is excellent at shock capturing.

\subsection{Collisions}~\label{subsec:collsions}

We start by conducting the 1D dust-gas collision test as  a benchmark, similar to Section 3.1 of~\cite{Benitez2019FARGO3D}.
We consider two dust species with constant stopping time $T_{s,1}, T_{s,2}$, and set three collision tests named A, B and C. The gas and all dust species are homogeneous, each having its own density ($\rho_\tx{g}, \rho_{\tx{d},1}, \rho_{\tx{d},2}$) and velocity ($v_\tx{g}, v_{\tx{d},1}, v_{\tx{d},2}$). The system then evolves under the mutual aerodynamic drag forces, characterized by two eigenvalues $\lambda_1$, $\lambda_2$, in the form of
\begin{equation}\label{eq:collision}
    v = v_\tx{COM} + c_1 \exp{(\lambda_1\;t)} + c_2 \exp{(\lambda_2\;t)}\ ,
\end{equation}
where $v_\tx{COM}$ is the center-of-mass velocity of the system. Their initial conditions, as well as the associated coefficients and eigenvalues are given by shown in Table~\ref{tab:collision}, and we provide the calculation procedures in Appendix~\ref{app:ana_collision}. The three tests are designed to test the non-stiff case (Test A), the stiff case with small stopping time (Test B) and the stiff case with large dust-to-gas ratios (Test C).
These tests are conducted in 1D Cartesian coordinates with a periodic boundary condition. We use the adiabatic equation of state with the adiabatic index being $\gamma = 1.4$ and an initial gas sound speed is set as $c_\tx{s}^2 \equiv \gamma \frac{P}{\rho_\tx{g}} = 1.4$ for all three tests. We include the work and friction heating from drags in the energy equation. We have tested eight drag integrators (Explicit: ``RK1-Explicit'', ``RK2-Explicit'', ``VL2-Explicit''; Semi-Implicit: ``Trapezoid'', ``TrBDF2''; Fully-Implicit: ``RK1-Implicit'', ``RK2-Implicit'' and ``VL2-Implicit''), and the main results are discussed below.

\begin{table}[htp]
\begin{tiny}
\centering
\caption{The Initial Conditions, Eigenvalues and Coefficients of the Analytic Solutions in the Collision Tests}
\begin{tabular}{cccc}
\toprule
Test            & A                 & B                 & C                 \\
\hline
$\rho_\tx{g}$   & 1                 & 1                 & 1                 \\
$v_\tx{g}$      & 1                 & 1                 & 1                 \\
$\rho_\tx{d,1}$ & 1                 & 1                 & 10                \\
$v_\tx{d,1}$    & 2                 & 2                 & 2                 \\
$T_\tx{s,1}$    & 2                 & 0.01              & 2                 \\
$\rho_\tx{d,2}$ & 1                 & 1                 & 100               \\
$v_\tx{d,2}$    & 0.5               & 0.5               & 0.5               \\
$T_\tx{s,2}$    & 1                 & 0.002             & 1                 \\
\hline
Coefficients\tnote{1}  &                   &                   & \\
$v_\tx{COM}$   & 1.16666666666667  & 1.16666666666667  & 0.63963963963963  \\
$\lambda_1$    & -0.63397459621556 & -141.742430504416 & -0.52370200744224 \\
$\lambda_2$    & -2.36602540378444 & -1058.25756949558 & -105.976297992557 \\
$c_\tx{g,1}$   & -0.22767090063074 & -0.35610569612832 & -0.06458203330249 \\
$c_\tx{g,2}$   & 0.06100423396407  & 0.18943902946166  & 0.42494239366285  \\
$c_\tx{d,1,1}$ & 0.84967936855889  & 0.85310244713865  & 1.36237475791577  \\
$c_\tx{d,1,2}$ & -0.01634603522555 & -0.01976911380532 & -0.00201439755542 \\
$c_\tx{d,2,1}$ & -0.62200846792815 & -0.49699675101033 & -0.13559165545855 \\
$c_\tx{d,2,2}$ & -0.04465819873852 & -0.16966991565634 & -0.00404798418109 \\
\bottomrule
\end{tabular}
\label{tab:collision}
\begin{tablenotes}
\footnotesize
\item[1] The analytic solutions of velocities of gas and dust are in the form of Equation (\ref{eq:collision}). The analytic solutions of gas energy can be obtained by integrating the right hand side of the Equation (\ref{eq:gas_erg}) with $\omega = 1$.
\end{tablenotes}
\end{tiny}
\end{table}

\begin{figure*}[htp]
\centering
\includegraphics[scale=0.40]{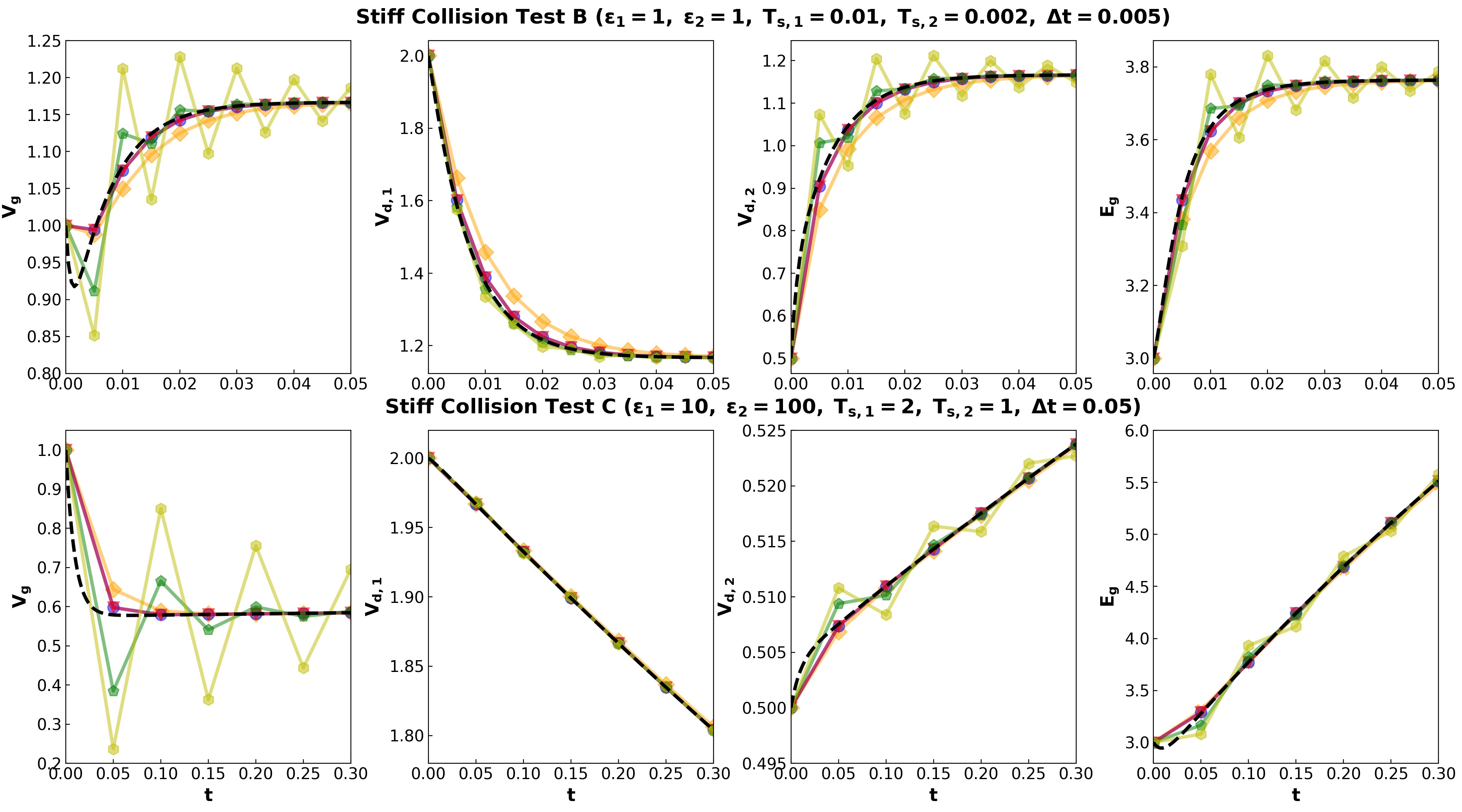}
\includegraphics[scale=0.40]{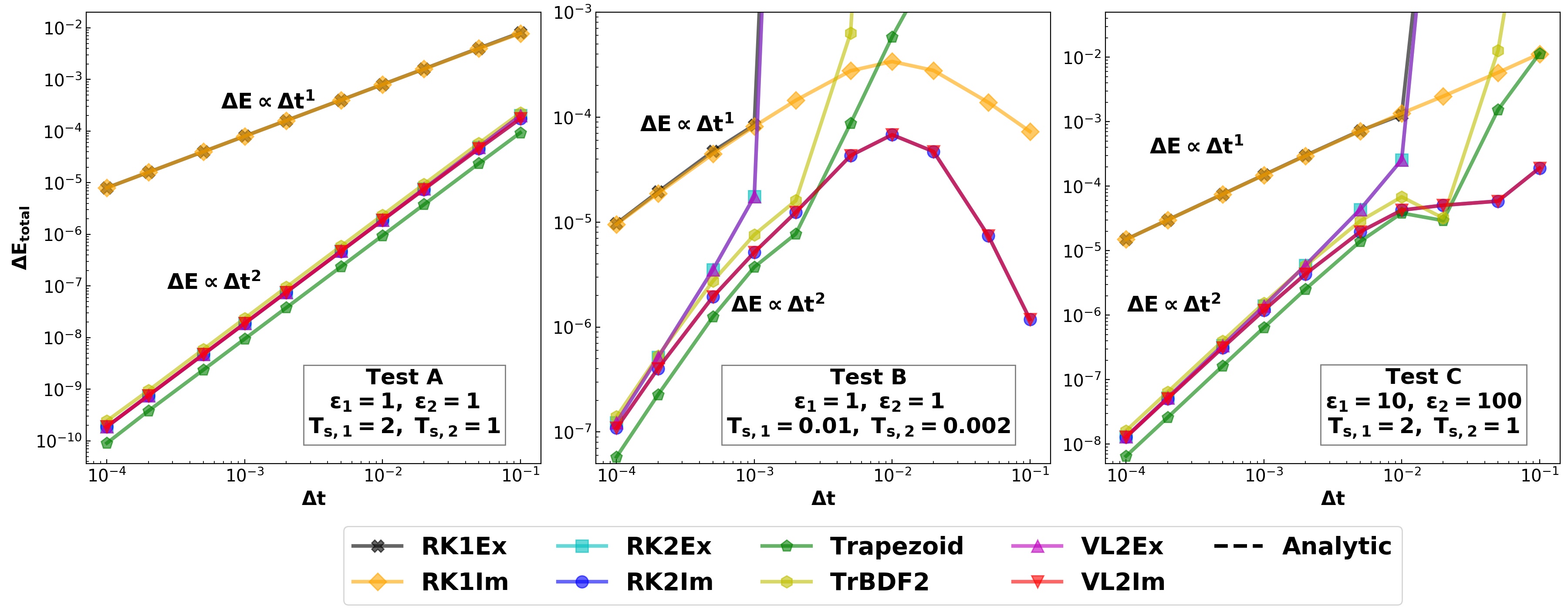}
\caption{The top eight panels are the velocities and energy of gas and dust in the collision Test B and C with the numerical time step $\Delta t = 0.005$ and $\Delta t = 0.05$. Solutions obtained by semi-implicit and fully-implicit different drag integrators are shown. The black dashed lines represent the analytic solutions. Note that explicit methods are unstable for our choice of $\Delta t>1/|\lambda_2|$ (see Table~\ref{tab:collision}), and the results are not shown. The bottom three panels are the scaling of total relative error $\Delta E_\tx{total}$ with numerical time step $\Delta t$ for different drag integrators (see legends) in the collision Tests A (left), B (middle) and C (right).}
\label{fig:Momentum_Dust_Error}
\end{figure*}

The top eight panels of Figure~\ref{fig:Momentum_Dust_Error} shows the temporal evolution of velocities and energy of both gas and dust in Test B and C with numerical time steps $\Delta t = 0.005$ and $0.05$, respectively. The results are to be compared with the analytic solution. Table~\ref{tab:collision} shows the largest eigenvalue $|\lambda_2| \simeq 1058$ and $106$ for Test B and C. Therefore, the drag terms become stiff when $\Delta t > 1/|\lambda_2| \simeq 0.001$ and $0.01$ for B and C. The explicit integrators are unstable in Test B and C with $\Delta t = 0.005$ and $0.05$. The semi-implicit methods are stable in the stiff drags, however, the numerical updates oscillate around the analytic solutions artificially, which is not unexpected as similar behavior was observed in \citet{Bai2010particle}.
Our fully-implicit methods ``RK1-Implicit'', ``VL2-Implicit'' and ``RK2-Implicit'' handle these two stiff regimes (small stopping times and large dust-to-gas ratios) very well, and the two second-order integrators are clearly seen to be more accurate.

To test numerical convergence, we calculate the relative error $\Delta E$ as a function of the numerical time step $\Delta t$ with different drag integrators. The $\Delta E$ is calculate by:
\begin{equation}
  \Delta E \left(\Delta t\right) =  \frac{1}{t_\tx{max} - t_\tx{min}}\sum \frac{|U_{\tx{num}} \left(\Delta t\right) - U_{\tx{ana}}|}{U_{\tx{ana}}}\Delta t
\label{eq:time_convergenece1}
\end{equation}
where $U$ represents momentum or the gas energy, subscripts ``\tx{num}'' and ``\tx{ana}'' represent the numerical and analytic solutions, respectively. The scaling of the total relative error $\Delta E_\tx{total} = \Delta E_\tx{mom} + \Delta E_\tx{erg}$ with different time steps $\Delta t$ is shown in the bottom three panels of Figure~\ref{fig:Momentum_Dust_Error} for different drag integrators. We vary $\Delta t$ from $10^{-4}$ to $10^{-1}$ which are non-stiff for Test A, but get increasingly stiff for Test B and C. The total relative errors are calculated between $t_\tx{min} = 0$ and $t_\tx{max} = 10$.

We see that in the non-stiff regime of Test A, all drag integrators achieve first or second-order accuracy in time as desired, and there is no significant difference between integrators of the same order. In the stiff regime of Test B and C, we see that the error in explicit and semi-implicit integrators diverge when $\Delta t\gtrsim(1/|\lambda_2|)$. The threshold for error divergence is only slightly higher for semi-implicit integrators. The fully-implicit integrators, on the other hand, achieve the desired level of accuracy at small $\Delta t$, while remain stable at large $\Delta t$ for both tests. Among them, the 2nd-order fully-implicit integrators show error levels that are at least one order of magnitude smaller, and are the preferred choices that we generally recommend.

Our drag integrators conserve total momentum of the dust-gas system by construction. While not shown in the figures, we have verified that in all three tests, total momentum is conserved to the fractional level of $\lesssim10^{-14}$ (i.e., approaching machine precision) within the duration of the simulations.

\subsection{Momentum Correction in Dust Diffusion}\label{subsec:momentum_correction}

Novel additions in our dust concentration diffusion formulation include the time derivative of the $\rho_\tx{d}\mbs{v}_{\tx{d,dif}}$ term and the $\mbs{\Pi}_\tx{d,dif}$ in the dust momentum equation (\ref{eq:dust_mom}). We refer to them as ``momentum correction''. They reflect the fact that the concentration diffusion flux carries momentum that back-react to the gas, and that the new formulation is Galilean invariant. To test these aspects of our implementation, we consider the following simple test problems without/with momentum correction in 1D, 1.5D and 2D Cartesian coordinates, where $1.5D$ means 1D test in the presence of a transverse velocity.

\subsubsection{Initial Setups}

We give the parameters of our tests in Table~\ref{tab:diffusion}, with more details below.
The 1D and 1.5D tests are carried in Cartesian coordinates, with 256 uniform grids, covering $x \in [0, 20]$. The gas has uniform initial density $\rho_{\tx{g0}} = 1$ and a single dust species with constant stopping time $T_\tx{s} = 10^{-2}$ is included with an initial Gaussian density profile:
\begin{equation}
  \rho_{\tx{d0}} = A \exp{\left[-\frac{\left(x-x_0\right)^2}{2\sigma_x^2}\right]} + \rho_{\tx{g0}}\ .
\end{equation}
where $A = 5$, $x_0 = 10$ and $\sigma_{x} = 2$ in both cases. We include gas viscosity and dust diffusion with coefficients $\nu = D_\tx{d} = 1$. In the 1D test, initial gas and dust velocities are zero, whereas in the 1.5D test, gas and dust have constant transverse velocities $v_{\tx{g},y} = v_{\tx{d},y} = 1$.

In the 2D test, we use 2D Cartesian coordinates in the domain $x, y \in[0, 20]$ with $256^2$ cells, and set the initial 2D Gaussian dust density profile in the center of the box:
\begin{equation}
  \rho_{\tx{d0,2D}} = A \exp{\left[- \frac{\left(x-x_0\right)^2}{2\sigma_{x}^2} - \frac{\left(y-y_0\right)^2}{2\sigma_{y}^2}\right]} + \rho_{\tx{g0}}\ .
\end{equation}
where $y_0 = 10$ and $\sigma_{y} = 2$, and the rest of parameters are same as in the 1.5D test.

We note that when including momentum correction, there is an additional contribution $v_\tx{d,dif}$ to the conserved dust momentum. This leads to two possible initial settings. One is to make the initial conserved momentum to be zero. By the conversion relation (\ref{eq:dustvar}), the primitive velocity thus equals to $-v_{\tx{d,dif}}$. Alternatively, one can choose the primitive velocity to be zero (i.e., zero mean dust velocity), so that the conserved momentum becomes $\rho v_{\tx{d,dif}}$ (i.e., non-zero mean dust momentum).
This ambiguity reflects that the initial condition itself is physically unrealistic to build up without involving additional source terms. As a test problem, we choose the latter, which we consider to be physically more natural and intuitive (the alternative choice would lead to different interpretable outcomes that we omit here for brevity). Note that without momentum correction, we set $v_{\tx{d}}=0$ so that the two setups share the same initial condition in primitive variables.

\begin{table}[]
\centering
\begin{footnotesize}
\caption{Problem setup for 1D, 1.5D and 2D dust diffusion tests}
\begin{tabular}{cccccc}
\toprule
\multicolumn{1}{c}{}  & Correction\footnote{The correction terms by default include both $\mbs{\Pi}_\tx{dif}$ and $\pt{(\rho_\tx{d} \mbs{v}_{\tx{d,dif}})}/\pt{t}$.} & $v_{\tx{g},x}$ & $v_{\tx{g},y}$ & $v_{\tx{d},x}$ & $v_{\tx{d},y}$ \\
\hline
\multirow{2}{*}{1D}   & No         & 0              & 0              & 0              & 0              \\
& Yes        & 0              & 0              & $v_{\tx{d,dif},x}$\footnote{The diffusion velocity $v_\tx{d,dif}$ is calculated by Equation (\ref{eq:conflx}).}          & 0              \\
\hline
\multirow{2}{*}{1.5D} & No         & 0              & 1              & 0              & 1              \\
& Yes        & 0              & 1              & $v_{\tx{d,dif},x}$          & 1              \\
\hline
\multirow{2}{*}{2D}   & No         & 1              & 1              & 1              & 1              \\
& Yes\footnote{In the 2D cases, we have two corrections tests, one is with only $\mbs{\Pi}_\tx{dif}$, and the other is with both $\mbs{\Pi}_\tx{dif}$ and $\pt{(\rho_\tx{d} \mbs{v}_{\tx{d,dif}})}/\pt{t}$.}        & 1              & 1              & 1+$v_{\tx{d,dif},x,\tx{2D}}$       & 1+$v_{\tx{d,dif},y,\tx{2D}}$       \\
\bottomrule
\end{tabular}
\label{tab:diffusion}
\end{footnotesize}
\end{table}

In all these tests, dust back-reaction is included. We use an isothermal equation of state with sound speed $c_\tx{s,iso} = 1$, applying periodic boundary conditions. We use the Piecewise Linear Method (PLM) spatial reconstruction for both gas and dust, and a CFL number of 0.3. The ``VL2-Implicit'' drag integrator is used to calculate the mutual drags.

\subsubsection{Results}

\begin{figure*}[htp]
\centering
\includegraphics[scale=0.40]{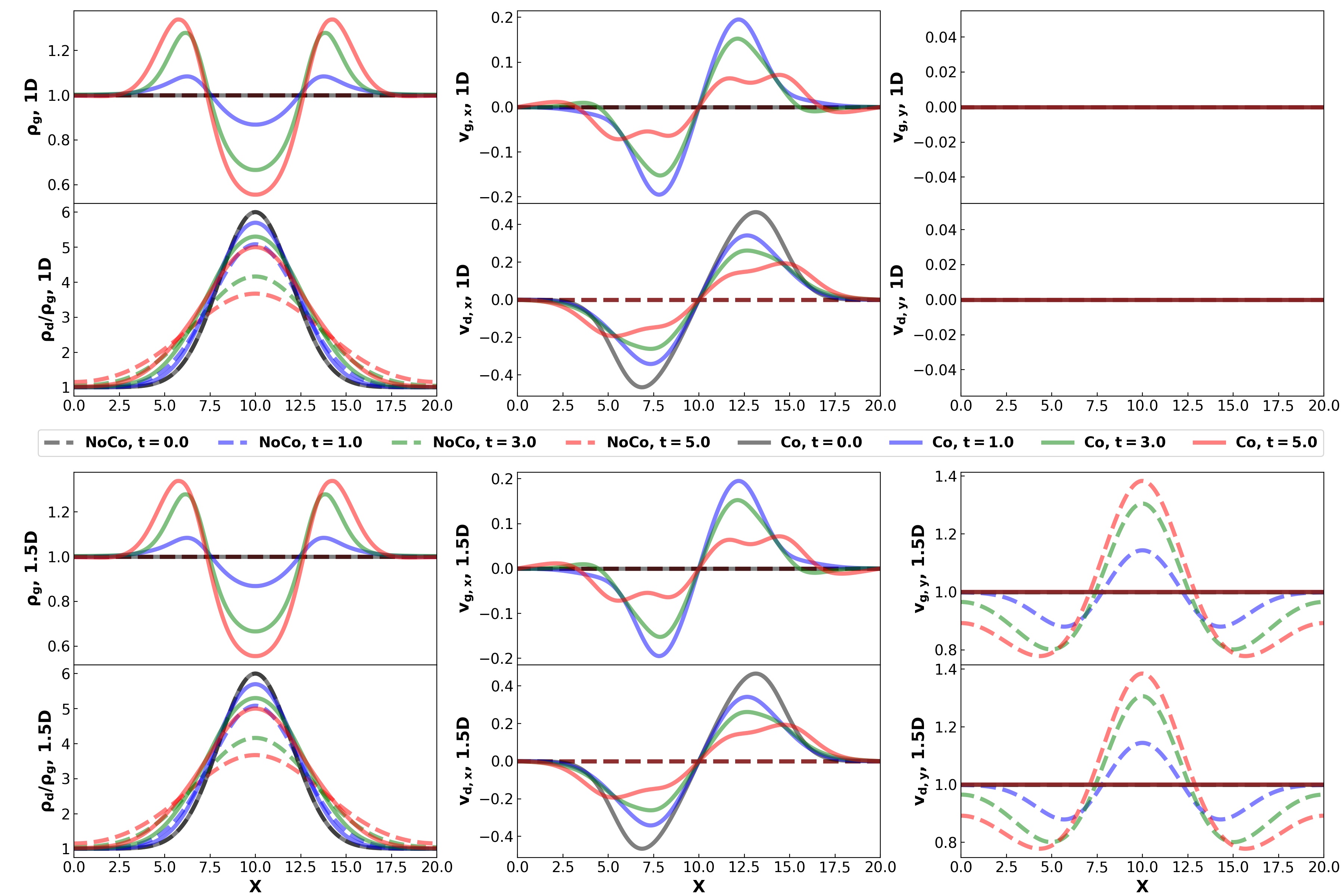}
\caption{1D (top six panels) and 1.5D (bottom six panels) dust diffusion tests without (``No Correction, NoCo'', dashed lines)/with (``Correction, Co'', solid lines) momentum corrections. The first and the third rows show gas density ($\rho_\tx{g}$) and velocities ($v_{\tx{g},x},\; v_{\tx{g},y}$) at different times ($t = 0.0,\;1.0,\;3.0$ and $5.0$). Note that the gas density and $x$-velocity in the first and third panels are identically one and zero in the ``NoCo'' cases with these dashed lines embedded below the black solid line. The second and the fourth rows show dust concentration ($\rho_\tx{d}/\rho_\tx{g}$) and dust velocities ($v_{\tx{d},x},\; v_{\tx{d},y}$) at different times. Similarly, the dust $x$-velocity in the ``NoCo'' case is zero with the corresponding dashed lines embedded below the black solid line.}
\label{fig:1D_Momentum_Diffusion}
\end{figure*}

\begin{figure*}[htp]
\centering
\includegraphics[scale=0.40]{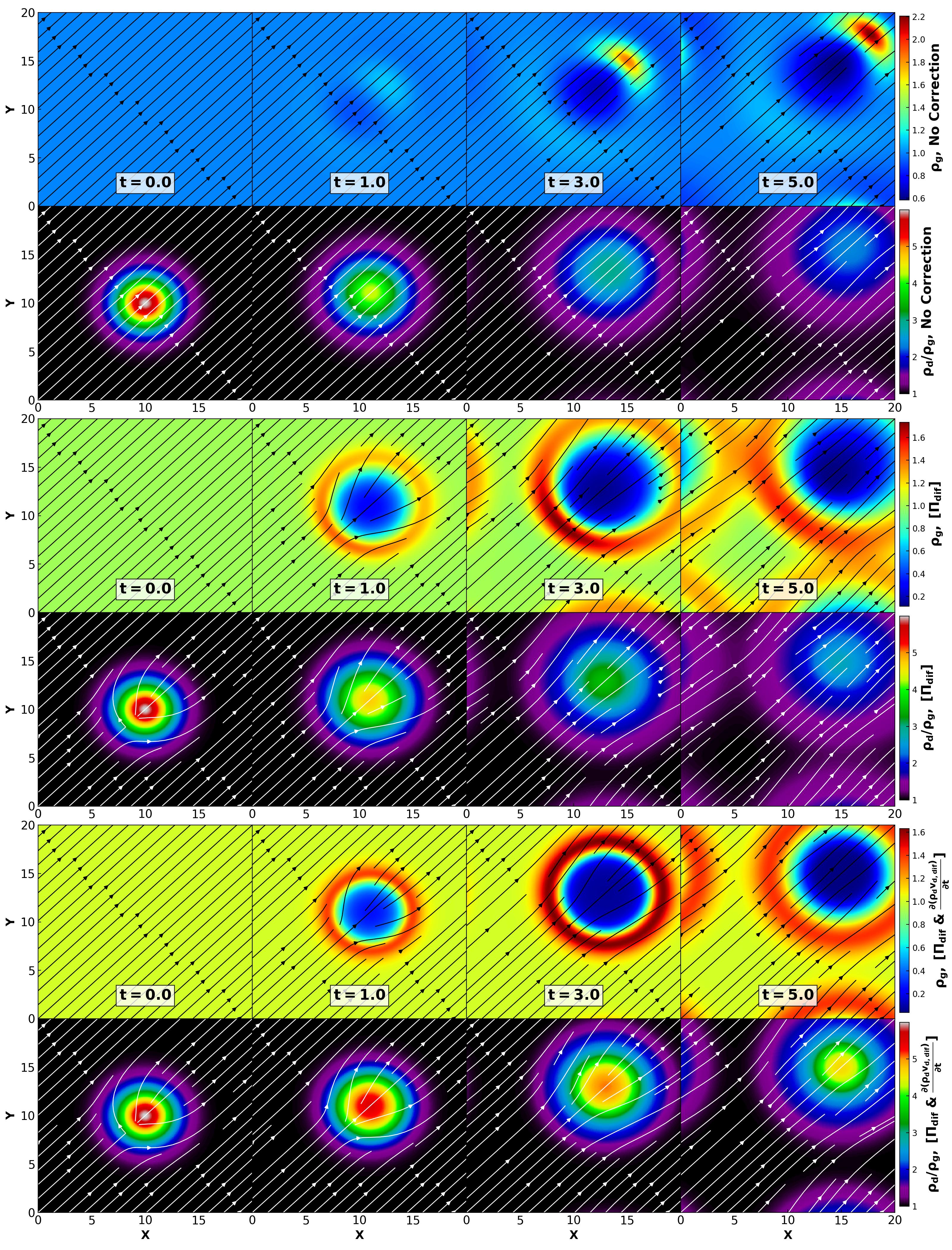}
\caption{2D dust diffusion tests without correction (top eight panels), with only $\mbs{\Pi}_\tx{dif}$ correction (middle eight panels), and with both $\mbs{\Pi}_\tx{dif}$ and $\pt (\rho_\tx{d} \mbs{v}_\tx{d,dif})/\pt t$ corrections (bottom eight panels). From left to right, the panels are at time $t = 0.0, \;1.0, \;3.0$ and $5.0$. The first, the third and the fifth rows are for gas density ($\rho_\tx{g}$), while the second, the fourth and the sixth rows are for dust concentration ($\rho_\tx{d}/\rho_\tx{g}$). The black (white) lines represent the velocity streamlines of gas (dust).}
\label{fig:2D_Momentum_Diffusion}
\end{figure*}

In the 1D tests shown in the top eight panels of Figure~\ref{fig:1D_Momentum_Diffusion}, we note that the correction term $\mbs{\Pi}_{\tx{dif}}$ is zero, thus only the time derivative term $\frac{\pt{(\rho_\tx{d} \mbs{v}_{\tx{d,dif}})}}{\pt{t}}$ is effective. When the correction is not included, dust proceeds as normal concentration diffusion, whereas the gas is totally intact. This is unphysical because turbulent mixing is a two-way process that not only mixes dust with gas, but should also mix gas with dust. When the momentum correction is included, we see that gas density exhibits a central deficit and two outside bumps. This is essentially the outcome of gas being dragged by the outward diffusion flow of dust, as can be seen in the central region of the middle panel. The additional structures in the gas act to slow down dust concentration diffusion, and we can see that without incorporating momentum correction, dust diffuses more than 2 times more rapidly.

When adding a transverse velocity in the 1.5D test, one should expect identical results as in 1D except for a velocity shift in the $y$-direction. However, without momentum correction, we see in the two right panels in the bottom of Figure~\ref{fig:1D_Momentum_Diffusion} that the system develops artificial variations in $v_y$. This is because concentration diffusion changes the dust density profile, but without properly altering the momentum profile. When momentum corrections are included, we see that dust and gas momenta are properly advected to ensure Galilean invariance.

In the 2D tests shown in Figure~\ref{fig:2D_Momentum_Diffusion}, similar to the 1D case, incorporating momentum correction leads to the physical behavior where the gas density develops a dip in the center surrounded by an outside ring, which would be absent without the correction. To test Galilean invariance, we add a background advection velocity along the diagonal direction. We run two cases with momentum correction, one includes only the $\mbs{\Pi}_\tx{dif}$ term, and the other includes both the $\mbs{\Pi}_\tx{dif}$ and $\pt (\rho_\tx{d} \mbs{v}_\tx{d,dif})/\pt t$ terms. We see that the circular gas and dust density patterns are well preserved when both terms are included, whereas artificial density structures show up without the correction or when including only the $\Pi_\tx{dif}$ term. These artificial density patterns are due to artifacts in velocity variations resulting from the violation of Galilean invariance. Therefore, both correction terms are essential to ensure Galilean invariance. Overall, these tests demonstrate the importance of properly incorporating the momentum correction terms to for a consistent treatment of dust concentration diffusion, especially when dust feedback is taken into account.

\subsection{Streaming Instability}\label{subsec:SI}

The streaming instability (SI) is a stringent test for two-way gas drag, and its nonlinear evolution with strong dust clumping represents a further test of code capability to handle sharp discontinuities. We adopt the linear tests given by~\cite{YoudinJohansen2007} and non-linear runs from~\cite{JohansenYoudin2007}, which has become the standard test problem for codes with particle-based treatment of dust ~\citep{Bai2010particle}, and more recently for multifluid dust as well~\citep{Benitez2019FARGO3D}. We will further extend the non-linear tests to cylindrical coordinates (Section~\ref{subsec:RunBA_global}), and to incorporate mesh refinement (Section~\ref{subsubsec:RunAB_SMR}).

\subsubsection{Shearing Box Equations and Equilibrium State}\label{subsubsec:shearing}

Most of our SI tests are carried out in the local shearing box framework, which follows a local patch of a disk at some fiducial radius $r_0$ in the corotating frame with orbital frequency $\Omega_0 = \Omega\left(r_0\right)$~\citep{GoldreichLyndenBell1965,HawleyGammie1995}. The equations of gas and dust are written in a Cartesian coordinate system $\left(x,y,z\right)$ for radial, azimuthal and vertical directions with Coriolis and centrifugal source terms~\citep{Stone2010}. The unit vectors along this three directions are denoted as $\left(\hat{i}, \hat{j}, \hat{k}\right)$, respectively. We do not consider viscosity, diffusion, magnetic field and self-/vertical gravity. Adopting an isothermal equation of state with isothermal sound speed $c_\tx{s}$, the continuity and momentum equations of gas and dust are:
\begin{equation}
\begin{aligned}
\frac{\pt \rho_{\tx{g}}}{\pt t} + \nabla \cdot \left(\rho_{\tx{g}} \mbs{v}_{\tx{g}}\right) =0\ ,
\end{aligned}
\end{equation}

\begin{equation}
\begin{aligned}
&\frac{\pt \left(\rho_{\tx{g}} \mbs{v}_{\tx{g}}\right)}{\pt t} + \nabla \cdot \left(\rho_{\tx{g}} \mbs{v}_{\tx{g}} \mbs{v}_{\tx{g}}+ P_{\tx{g}} \mathsf{I}\right)=  \\
2 \rho_{\tx{g}} q \Omega_{0}^{2 } x \hat{\mbf{i}} & -2 \rho_{\tx{g}} \Omega_{0} \hat{\mbf{k}} \times \mbs{v}_{\tx{g}} - \sum^{n}_{k=1}\rho_{\tx{d},k} \frac{\mbs{v}_{\tx{g}} - \mbs{v}_{\tx{d},k}}{T_{\tx{s},k}}\ ,
\end{aligned}
\end{equation}

\begin{equation}
\begin{aligned}
\frac{\pt \rho_{\tx{d},k}}{\pt t} +\nabla \cdot \left(\rho_{\tx{d},k} \mbs{v}_{\tx{d},k} \right) = 0\ ,
\end{aligned}
\end{equation}

\begin{equation}
\begin{aligned}
\frac{\pt \left(\rho_{\tx{d},k} \mbs{v}_{\tx{d},k}\right)}{\pt t}+ \nabla \cdot &\left(\rho_{\tx{d},k} \mbs{v}_{\tx{d},k} \mbs{v}_{\tx{d},k}\right)= \\
2 \rho_{\tx{d},k} q \Omega_{0}^{2} x \hat{\mbf{i}} -2 \rho_{\tx{d},k} \Omega_{0} \hat{\mbf{k}} &\times \mbs{v}_{\tx{d},k} + \rho_{\tx{d},k}\frac{\mbs{v}_{\tx{g}} - \mbs{v}_{\tx{d},k}}{T_{\tx{s},k}}\ .
\end{aligned}
\end{equation}
where 
$q=-d\ln{\Omega}/dr$ is the shear rate and $q = 3/2$ for Keplerian disks as we adopt.

Ignoring vertical gravity (i.e., unstratified disk), dust and gas can achieve the so-called Nakagawa-Sekiya-Hayashi (NSH) equilibrium~\cite[]{NakagawaSekiyaHayashi1986}. The equilibrium is a force balance between background pressure gradient, centrifugal force, Coriolis force and mutual aerodynamic drags between gas and dust in the horizontal plane. Here we consider a Keplerian disk with angular speed $\Omega_0=\Omega_\tx{K}$. In the absent of dust, gas rotates slower than the Keplerian speed by a small amount $\eta v_K\equiv\Pi c_\tx{s}$ due to the background pressure gradient, and $\Pi$ is $\lesssim0.1$ under typical disk conditions.  $\eta$ represents the strength of radial gas pressure gradient:
\begin{equation}
  \eta \equiv  \frac{1}{2} \frac{\ln P_\tx{g}}{\ln r} \left(\frac{h}{r}\right)^2 =  \frac{1}{2} \frac{\ln P_\tx{g}}{\ln r} \left(\frac{c_\tx{s}}{v_\tx{k}}\right)^2 \ ,
\label{eq:eta}
\end{equation}
The original NSH solution considered a single dust species. Generalized to multiple dust species with different stopping times, the velocities of gas and dust in the multi-species equilibrium are given by~\citep{Benitez2019FARGO3D}, and also see~\cite{TanakaHimenoIda2005}:
\begin{equation}
\begin{aligned}
  v_{\tx{g0},x}   &= 2 \eta v_\tx{K} \frac{\mathcal{A}}{\mathcal{A} +\mathcal{B}}\ ,\quad
  v_{\tx{d0},k,x} = \frac{v_{\tx{g0},x}+2 St_{k} v'_{\tx{g0},y}}{1+ St_{k}^{2}}\ , \\
  v'_{\tx{g0},y}   &= - \eta v_\tx{K} \frac{\mathcal{B}}{\mathcal{A} +\mathcal{B}}\ ,\quad
  v'_{\tx{d0},k,y} = \frac{v'_{\tx{g0},y}-St_{k} v_{\tx{g0},x}/2}{1+St_{k}^{2}}\ ,
\end{aligned}
\label{eq:NSH}
\end{equation}
where $St_\tx{k}\equiv\Omega_\tx{K}T_{\tx{s},k}$ is the dimensionless stopping time of k-th dust species, the $y-$ velocities with a prime have Keplerian shear subtracted $v'_{\tx{g}0,y}=v_{\tx{g}0,y}+(3/2)\Omega_Kx$, $v'_{\tx{d}0,k,y}=v_{\tx{d}0,k,y}+(3/2)\Omega_Kx$, and
\begin{equation}
\mathcal{A} =\sum_{k=1}^{n} \frac{\epsilon_{k} St_{k}}{1+ St_{k}^{2}}\ , \quad
\mathcal{B} =1+\sum_{k=1}^{n} \frac{\epsilon_{k}}{1+ St_{k}^{2}}\ .
\end{equation}

In our numerical setup, we add an additional outward force $\mbs{f} \equiv 2 \rho_\tx{g} \eta v_\tx{K} \Omega_0 \hat{i}$ on the gas to mimic the radial pressure gradient. Note that it differs from~\cite{Bai2010particle} and~\cite{Benitez2019FARGO3D}, who add this force on the dust component. The two approaches are equivalent except for a constant velocity shift. Realize the exact analytic solution of the multi-species NSH equilibrium is straightforward when using explicit integrators, and our numerical implementation in Section~\ref{sssec:combine} ensures that such exact solution can be realized using the semi-implicit and implicit drag integrators as well.

\subsubsection{Linear SI Modes and Growth Rates}\label{subsubsec:SI-Linear}

The NSH equilibrium is subject to the SI~\citep{YoudinGoodman2005}. The growth rate $s$ of the SI is a function of initial dust-to-gas ratio (or metallicity) $\epsilon_0 \equiv \rho_\tx{d,0}/\rho_\tx{g,0}$, dimensionless stopping time $St$ and two dimensionless wavenumbers $K_x \equiv k_x \eta r_0 = k_x \eta c_\tx{s}/\Omega_\tx{K} $ and $K_z \equiv k_z \eta r_0= k_z \eta c_\tx{s}/\Omega_\tx{K} $: $s = s(\epsilon_0, St, K_x, K_z)$. In this work, we choose the Lin-A and Lin-B tests in Table 1 of~\cite{YoudinJohansen2007}, which consist of gas and one dust species, and the Lin-3 test in Section 3.5 of~\cite{Benitez2019FARGO3D}, which consists of gas and two dust species.

The numerical setups of the linear tests are similar to those of~\cite{YoudinJohansen2007},~\cite{Bai2010particle} and~\cite{Benitez2019FARGO3D}. The numerical domain is a square box along $x$ and $z$ directions with $-L_x/2 \leq x \leq L_x/2$, $-L_z/2 \leq z \leq L_z/2$, and $L_x = L_z = 1$. We choose $\Omega_0 = 1.0$ and $\eta v_\tx{K} = 0.05\;c_\tx{s}$.
On top of the multi-species NSH equilibrium from Equation (\ref{eq:NSH}), we add perturbations on densities and velocities of both gas and dust of the following form, following~\citet{YoudinJohansen2007}:
\begin{equation}
\begin{aligned}
  \delta\rho &= [\Re(\tilde \rho) \cos \phi - \Im(\tilde \rho) \sin\phi] \cos (k_z z)\ ,\\
  \delta v_x &= [\Re(\tilde v_x) \cos\phi - \Im(\tilde v_x) \sin\phi] \cos (k_z z)\ , \\
  \delta v_y &= [\Re(\tilde v_y) \cos\phi - \Im(\tilde v_y) \sin\phi] \cos (k_z z)\ , \\
  \delta v_z &= [\Re(\tilde v_z) \sin\phi  + \Im(\tilde v_z) \cos\phi] \sin (k_z z)\ . \\
\end{aligned}
\end{equation}
where $\phi\equiv k_x x-\omega t$, and density and velocity perturbations, denoted by $\tilde\rho$, $\tilde{\boldsymbol v}$ for gas and individual dust species, are given by the respective eigenvectors of the specific linear modes, which are listed in Table 1 of~\cite{YoudinJohansen2007} and Table 4 of~\cite{Benitez2019FARGO3D}.

In these tests, we fit one eigenmode in the simulation box. As our box size is fixed, the sound speed $c_\tx{s}$ is no longer a free parameter and it depends on the dimensionless wavenumbers ($K_x$ and $K_z$). Because of $K_z = K_x = k_x \eta r_0 = 2\pi\times 0.05 c_\tx{s}/ \Omega_0$, we obtain $c_\tx{s} = K_x \Omega_0/(0.05 \times 2 \pi)$. In the Lin-A test, $K_x = K_z = 30$, thus we have $c_\tx{s,A} = 95.49296585$. Similar setup can be done in Lin-B and Lin-3 tests, with $c_\tx{s,B} = 19.09859317$ in Lin-B and $c_\tx{s,3} = 159.1549431$ in Lin-3. We use all 2nd-order drag integrators, with PLM and Piecewise Parabolic Method (PPM) for spatial reconstructions on both gas and dust, and the HLLE Riemann solver. Numerical resolution spans from 8 to 256 cells per wavelength in all each directions, with the CFL number being 0.3.

\begin{figure*}[htp]
\centering
\includegraphics[scale=0.40]{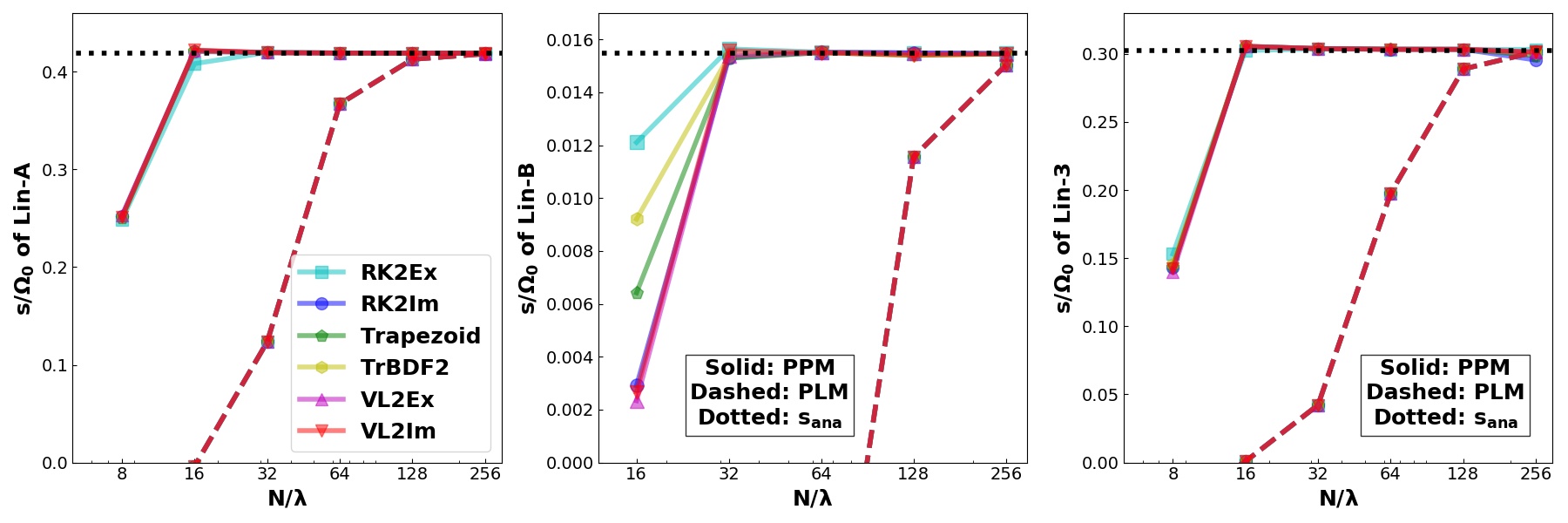}
\caption{Fitted averaging growth rates (normalized as $s/\Omega_0$) by measuring the mean kinetic energy $E_\tx{kinetic}$ of gas and dust in Lin-A, Lin-B and Lin-3 tests of SI. Results are shown as a function of the number of grids per wavelength ($N/\lambda$). The analytic growth rates are marked with blacked dotted lines. Different drag integrators are marked with different colors. The solid (dashed) lines represent the runs with PPM (PLM) spatial reconstruction.}
\label{fig:growth_rate_SI}
\end{figure*}

We measure the growth rate of the kinetic energy $E_\tx{kinetic}$ of both gas and dust in the linear tests. Because the initial kinetic energy is dominated by radial drift, we expect the temporal variation of kinetic energy to grow as $\delta E_\tx{kinetic} \propto \exp(s\;t)$.
We fit the spatial standard deviation $<\delta E_\tx{kinetic}^2>^{\frac{1}{2}}$ over the whole mesh as a function of time, and show the fitted growth rate $s/\Omega_0$ in Figure~\ref{fig:growth_rate_SI} as a function of resolution in Lin-A, Lin-B and Lin-3 tests. As these test problems are non-stiff, different drag integrators generally show similar results. With PPM reconstruction, 16 cells per wavelength is generally sufficient to accurately capture the growth in all three tests. When using the PLM reconstructions, on the other hand, about 128-256 cells per wavelength is needed for similar accuracy, and the requirement for Lin-B is the most stringent.
When compared with the measured growth rates in Table 5 of~\cite{Benitez2019FARGO3D}, it appears that we need more grid points to achieve similar accuracy in FARGO3D. On the other hand, the level of accuracy that we achieve is similar to those obtained in other finite volume method code, Athena~\citep{Bai2010particle} and PLUTO~\citep{MignoneFlock2019Dust}, with PPM spatial reconstructions. We thus attribute the difference primarily to the different nature of the base code. On the other hand, we will show that our code show similar outcomes in the nonlinear regime at a given resolution.

\subsubsection{The SI in the Non-Linear Regime}\label{subsubsec:SINonLinear}

\begin{figure*}[htp]
\centering
\includegraphics[scale=0.45]{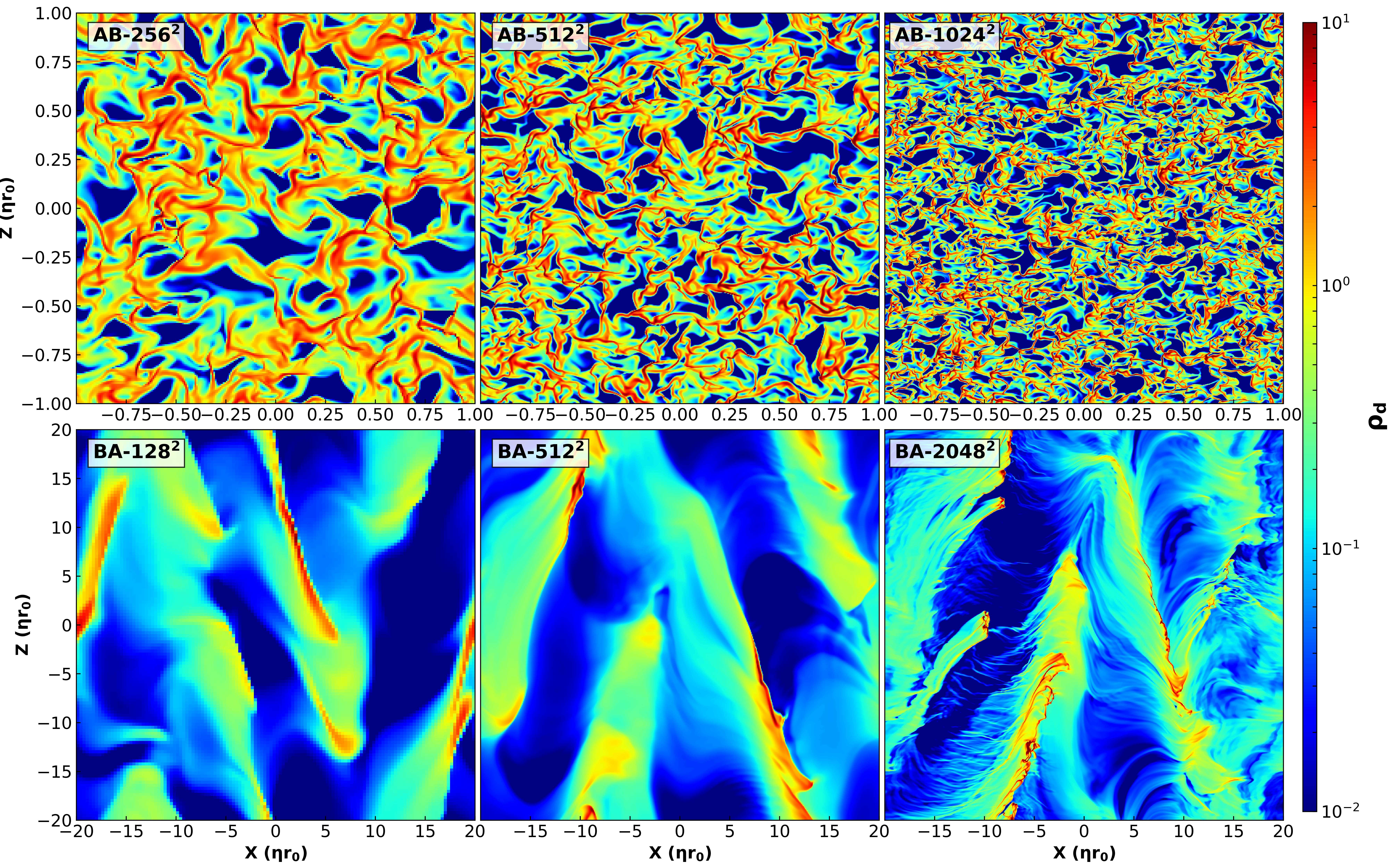}
\caption{Dust densities of AB test (top) and BA test (bottom) with different spatial resolutions at $t = 40\;\Omega_0^{-1}$ for AB test and $t=800\;\Omega_0^{-1}$ for BA test.
}
\label{fig:Nonlinear_SI}
\end{figure*}

\begin{figure*}[htp]
\centering
\includegraphics[scale=0.45]{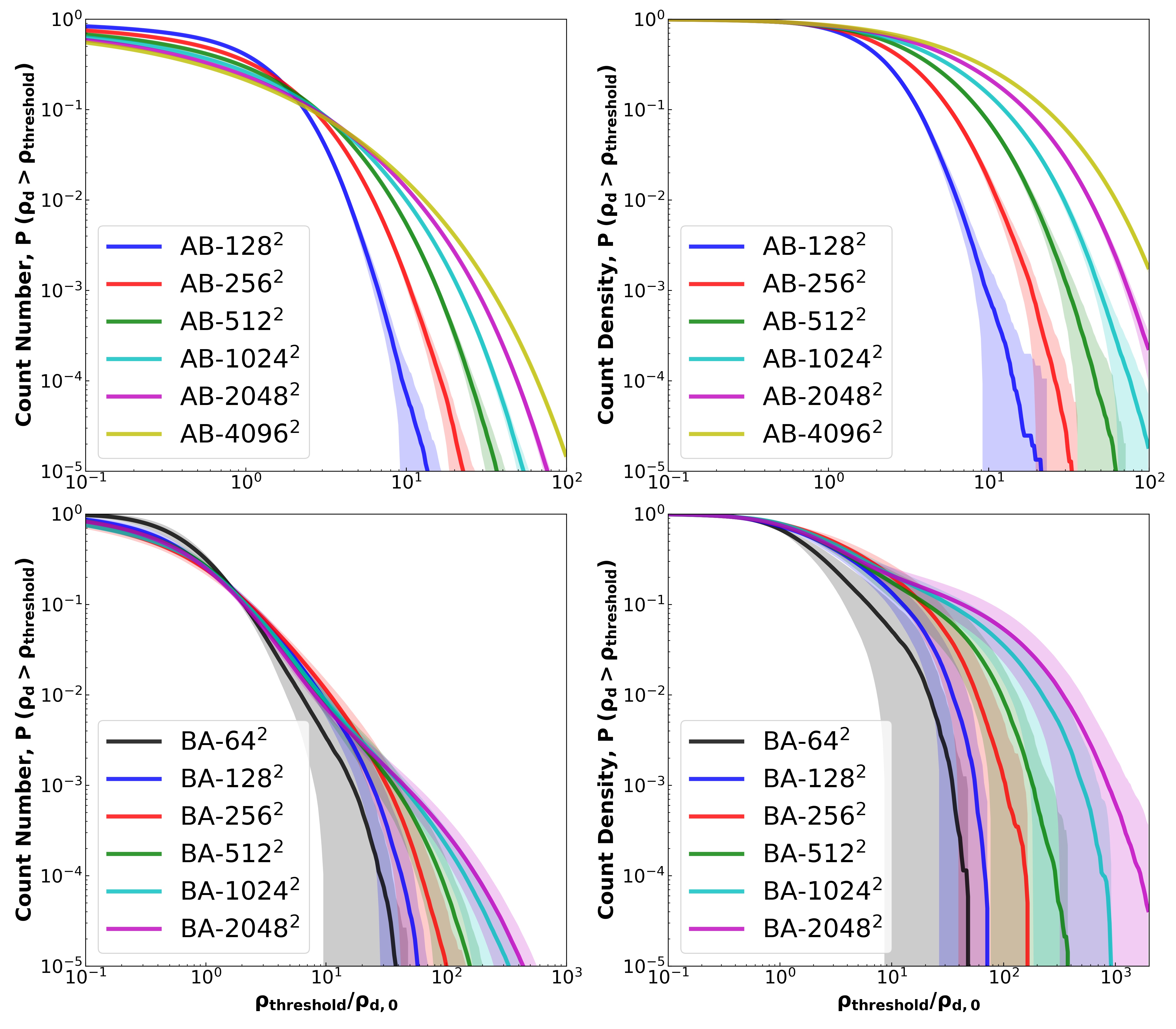}
\caption{The dust CDFs of AB test (top) and BA test (bottom), similar to Figure 10 of~\cite{Benitez2019FARGO3D}. The left panels are the CDFs calculated by counting the number of cells whose dust density exceeds certain threshold. The right panels are the CDFs calculated by additional weighting by dust density. The color shaded regions are the temporal standard deviations based on many snapshots (from $30\;\Omega_0^{-1}$ to $40\;\Omega_0^{-1}$ for AB test, and from $600\;\Omega_0^{-1}$ to $800\;\Omega_0^{-1}$ for BA test). The initial dust densities are $\rho_{\tx{d},0} = 1.0$ and $\rho_{\tx{d},0} = 0.2$ for AB and BA test, respectively.}
\label{fig:cumulative_density}
\end{figure*}

\begin{table*}[htp]
\begin{normalsize}
\centering
\caption{Turbulence Properties of AB and BA test with different resolutions}
\centering
\begin{tabular}{cccccc}
\toprule
Run & $Ma_x$ & $Ma_y$ & $Ma_z$ & $\mrs{Re}/\mrs{Re}_\tx{NSH}$ & $v_\tx{d,drift}/v_\tx{d,drift,NSH}$ \\
\hline
AB-$128^2$  & $1.39(06) \times 10^{-2}$ & $8.97(55) \times 10^{-3}$ & $1.18(08) \times 10^{-2}$ & $2.15(22)$ & $1.74(07)$ \\
AB-$256^2$  & $1.46(05) \times 10^{-2}$ & $9.30(49) \times 10^{-3}$ & $1.10(03) \times 10^{-2}$ & $2.56(10)$ & $2.03(08)$ \\
AB-$512^2$  & $1.37(05) \times 10^{-2}$ & $7.32(39) \times 10^{-3}$ & $1.07(06) \times 10^{-2}$ & $2.66(12)$ & $2.19(07)$ \\
AB-$1024^2$ & $1.24(02) \times 10^{-2}$ & $6.01(24) \times 10^{-3}$ & $8.95(28) \times 10^{-3}$ & $2.55(07)$ & $2.15(05)$ \\
AB-$2048^2$ & $1.10(02) \times 10^{-2}$ & $4.96(17) \times 10^{-3}$ & $7.61(15) \times 10^{-3}$ & $2.38(02)$ & $2.07(02)$ \\
AB-$4096^2$ & $1.07(01) \times 10^{-2}$ & $4.93(14) \times 10^{-3}$ & $7.26(14) \times 10^{-3}$ & $2.34(04)$ & $2.03(02)$ \\
\hline
BA-$64^2$   & $1.03(15) \times 10^{-2}$ & $1.69(14) \times 10^{-2}$ & $3.92(33) \times 10^{-2}$ & $0.74(08) $& $0.74(08)$\\
BA-$128^2$  & $1.46(23) \times 10^{-2}$ & $2.03(20) \times 10^{-2}$ & $4.92(26) \times 10^{-2}$ & $0.61(07) $& $0.63(06)$\\
BA-$256^2$  & $1.78(11) \times 10^{-2}$ & $2.09(15) \times 10^{-2}$ & $4.91(45) \times 10^{-2}$ & $0.54(08) $& $0.58(07)$\\
BA-$512^2$  & $1.59(21) \times 10^{-2}$ & $2.08(14) \times 10^{-2}$ & $5.02(16) \times 10^{-2}$ & $0.62(06) $& $0.65(05)$\\
BA-$1024^2$ & $1.57(15) \times 10^{-2}$ & $2.07(07) \times 10^{-2}$ & $4.83(13) \times 10^{-2}$ & $0.58(04) $& $0.62(04)$\\
BA-$2048^2$ & $1.61(15) \times 10^{-2}$ & $2.12(20) \times 10^{-2}$ & $4.95(24) \times 10^{-2}$ & $0.61(05) $& $0.64(05)$\\
\bottomrule
\end{tabular}
\label{tab:SI_turbulence}
\end{normalsize}
\begin{tablenotes}
\footnotesize
\item[1] The number in parenthesis quotes the $1\sigma$ uncertainty of the last two digits.
\end{tablenotes}
\end{table*}

The non-linear SI runs are also carried in the 2D shearing box in $x-z$.
We select two non-linear tests of the SI, namely, AB and BA test from~\cite{JohansenYoudin2007}. In AB test, the domain is $-\eta r_0 \leq x \leq \eta r_0 $, $-\eta r_0 \leq z \leq \eta r_0$, with parameters being $\epsilon_0 = 1.0$, $St = 0.1$. In BA test, the domain is $-20\;\eta r_0 \leq x \leq 20\;\eta r_0$, $-20\;\eta r_0 \leq z \leq 20\;\eta r_0$, with parameters being $\epsilon_0= 0.2$ and $St = 1.0$.
The parameters and simulation setups are similar to Table 1 of~\cite{JohansenYoudin2007}, Section 5 of~\cite{Bai2010particle} and Section 3.5.6 of~\cite{Benitez2019FARGO3D}.
We use the ``VL2-Implicit'' drag integrator, the Roe Riemann solver, with PPM reconstruction for the gas, and PLM reconstructions for the dust, in order to more robustly follow the dramatic variation of dust density over space in the non-linear stage with strong particle clumping.
We use an isothermal equation of state with sound speed is $c_\tx{s} = 1.0$, and the initial gas density is $\rho_\tx{g} = 1.0$. We also set a density floor $\rho_\tx{d,floor} = 10^{-6}$ on dust. Gas viscosity and dust diffusion are not included.
The simulations are initiated from the NSH equilibrium, on top of which we add white-noise velocity perturbations with an amplitude of $<A> \sim 0.02 c_\tx{s}$ on both gas and dust.

The saturated state of AB and BA tests, following the evolution of $40\;\Omega_0^{-1}$ and $800\;\Omega_0^{-1}$ are shown in Figure~\ref{fig:Nonlinear_SI}, for simulations with different resolutions. They are to be compared with with Figure 8 and 9 of~\cite{Benitez2019FARGO3D}.
AB test is characterized by the development of thin filaments and cavitation towards smaller scales at higher resolution, well consistent with results in previous works~\citep{JohansenYoudin2007,Bai2010particle,Benitez2019FARGO3D}.
In BA test, the system develops long dusty stripes and valleys nearly aligned with the $z$ direction and tilted towards the $x$ direction. Different from AB test, these general features are similar at all resolutions, again consistent with previous studies.

We further investigate the convergence of dust clumping by calculating the cumulative dust density distributions (CDFs). Following the same procedures described in Section 3.5.6 of~\cite{Benitez2019FARGO3D}, there are two ways of calculating the CDFs. One is based on the probability that local dust density exceeds certain threshold, obtained by counting the number of cells whose dust density exceeds the threshold. The other reflects the probability where a particle resides in regions whose particle density exceeds a certain threshold, obtained by weighting the first probability by local dust density. We refer to the two CDFs as obtained by counting cell numbers, and counting dust density, respectively. The overall results are shown in Figure~\ref{fig:cumulative_density}.

The CDFs obtained by counting cell numbers are very similar to those obtained by~\cite{Benitez2019FARGO3D} for both AB and BA. The CDFs of AB test systematically vary with resolution at both the low-density and high-density ends, in line with the non-convergent behaviors revealed in Figure~\ref{fig:Nonlinear_SI}. The CDFs of BA test show convergence at the low-density end up to $P\sim10^{-3}$, while show more significant clumping at higher resolution (BA-$1024^2$ and BA-$2048^2$).

Our CDFs by counting dust density, on the other hand, show more clumping than that in FARGO3D. The clumping is also more significant than that obtained in the particle module of Athena (see Figure 6 of \citealp{Bai2010particle}). This might be related to the higher-order drag integrators adopted here compared to FARGO3D. We also note that in the original Athena code, some artificial reduction of dust feedback was applied in strong dust clumps to alleviate the stiffness in the system that is circumvented in our approach. We thus leave this as an open issue. We also anticipate that our dust fluid module generally finds most of applications in regimes with $St<1$, instead of $St\gtrsim1$ as in BA test.

We also examine the properties of gas turbulence triggered by the SI. We calculate the turbulent Mach numbers along 3 directions ($x$, $y$ and $z$), the Reynolds stress and mean radial drift velocities of dust fluids. 
The mach number is calculated by $Ma \equiv \sqrt{\langle(v_\tx{g}-\ovl{v_\tx{g}})^2\rangle}/c_\tx{s}$. The Reynolds stress is calculated by: $ \mrs{Re} \equiv \langle\rho_\tx{g} v_{\tx{g},x} (v_{\tx{g},y} - v_\tx{K})\rangle$.
The mean radial drift velocity is computed by dividing the mean dust momentum over mean density.
These properties are all spatially and time averaged, indicated by angle brackets,
based on many snapshots ($30\;\Omega_0^{-1} \sim 40\;\Omega_0^{-1}$ in AB test, $600\;\Omega_0^{-1} \sim 800\;\Omega_0^{-1}$ in BA test), and the results are shown in Table~\ref{tab:SI_turbulence}. These diagnostic quantities are in broad agreement with the values obtained in~\cite{JohansenYoudin2007,Bai2010particle}, all saturated into highly subsonic, anisotropic turbulence, with enhanced radial drift and Reynolds stress in AB test and reduced radial drift and Reynolds stress in BA test.

\subsubsection{Global Curvilinear Run of BA test}\label{subsec:RunBA_global}

In order to test our multifluid module in curvilinear coordinates, we run a global unstratified SI test in  cylindrical coordinates $(r,\phi,z)$, similar to earlier investigations~\citep{Kowalik2013GlobalSI,MignoneFlock2019Dust}. We choose to adopt the parameters close to BA test here, which is less demanding in resolution and has better convergence properties. The computational domains in three directions are $r\in [0.2, 2.6]$, $\phi \in [0, 2\pi]$ and $z\in[-0.15, 0.15]$. The numerical resolutions are $4096\times 4 \times 512$ cells along $r$, $\phi$ and $z$, where we use a reduced $\phi-$resolution to preserve axisymmetry.

\begin{figure*}[htp]
\centering
\includegraphics[scale=0.3]{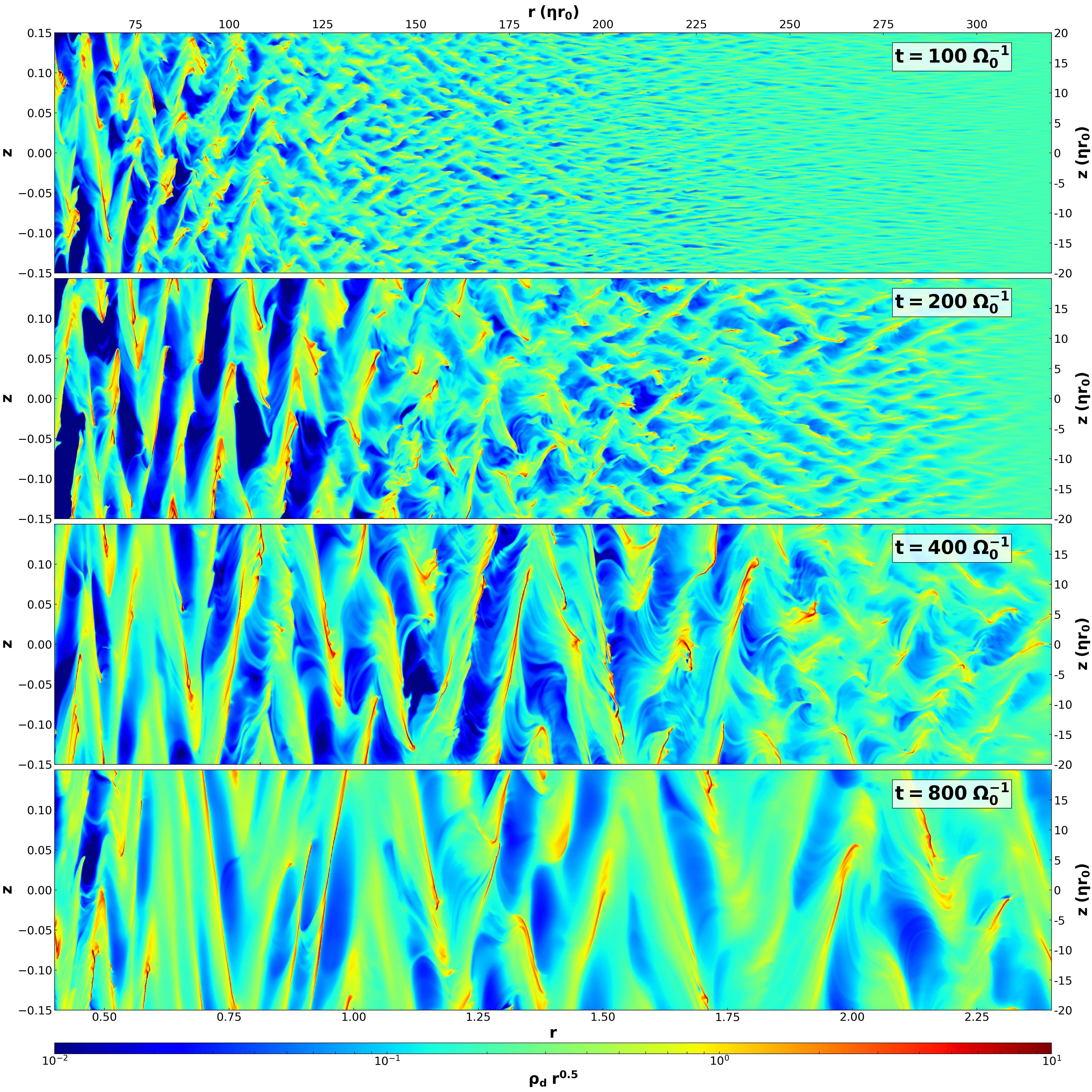}
\caption{Different snapshots of dust densities (scaled by $r^{-0.5}$) in the global unstratified BA test in the cylindrical coordinate. From top to bottom, the panels are at $t = 100\; \Omega_0^{-1}$, $200\;\Omega_0^{-1}$, $400\;\Omega_0^{-1}$ and $800\;\Omega_0^{-1}$, where $\Omega_0$ is the orbital frequency at $r_0 \equiv 1$. The strength of gas pressure gradient is $\eta = 0.0075$. The top and right axis are scaled by $\eta r_0$ for reference.
The effective resolution is the same as the BA-$512^2$ in Figure~\ref{fig:Nonlinear_SI}.}
\label{fig:Global_RunBA}
\end{figure*}

We set the central star mass $GM=1$, and set the gas radial density profile to be $\rho_\tx{g}(r) = \rho_0 (r/r_0)^{-0.5}$ with $\rho_0\equiv 1$ at $r_0\equiv 1$. We adopt a vertically-isothermal equation of state with $P(r) = c_\tx{s}(r)^2 \rho_\tx{g}(r)$. The sound speed is chosen so that the disk aspect ratio is $h/r=c_s/v_K=0.1$ at all radii, which gives $c_\tx{s} = 0.1(r/r_0)^{-0.5}$. From Equation (\ref{eq:eta}), we obtain $\eta = 0.0075$ ($\Pi=0.075$). The effective resolutions along $r$ and $z$ directions are $12.8/\eta r_0$, same as BA-$512^2$. The initial dust density follows that of the gas with a uniform dust-to-gas density ratio $\epsilon_0 = 0.2$. The Stokes number of dust is $St = T_\tx{s}\Omega_\tx{K} = 1.0$. The initial velocities of both gas and dust are set by the NSH equilibrium. Periodic boundary conditions are applied along $\phi$ and $z$ directions. The radial boundary condition is fixed by the NSH solution. To minimize of the unphysical wave reflection, we apply wave damping zones near the inner and outer radial boundaries~\citep{deVal-Borro2006,deVal-Borro2007}, located at $0.20 \leq r \leq 0.32$ and $2.44 \leq r \leq 2.60$, where gas and dust density and velocities (represented by $x$) are relaxed according to:
\begin{equation}
  \frac{\mathrm{d} x}{\mathrm{~d} t}=-\frac{x-x_\tx{init}}{\tau} R(r) \ ,
\end{equation}
where $x_\tx{init}$ is the initial value and $\tau$ is damping rate. We adopt $\tau = 1$ in our simulation, and $R(r)$ is a smoothing parabolic function that transiting from $0$ in the active zones to $1$ in the ghost zones. We also use the orbital advection algorithm~\cite[][FARGO algorithm]{Masset2000,Masset2002} to
reduce truncation error. At the beginning, we add small velocity perturbations in white noise with an amplitude of $0.02 c_\tx{s}$ to seed the instability.

Figure~\ref{fig:Global_RunBA} shows different snapshots of dust density in the global simulations of this global BA test. We can clearly see the progressive development of the SI from small radii to large radii, as the inner region has shorter dynamical time. The simulation reaches saturated state over the entire domain after about $t=600\;\Omega_0^{-1}$, where $\Omega_0=\Omega_\tx{K}$ at $r=r_0$, and we see the characteristic long dusty stripes and valleys with the maximum of $\rho_\tx{d} r^{0.5}$ significantly amplified by a factor of $\gtrsim100$.

\begin{figure*}[htbp]
\centering
\includegraphics[scale=0.45]{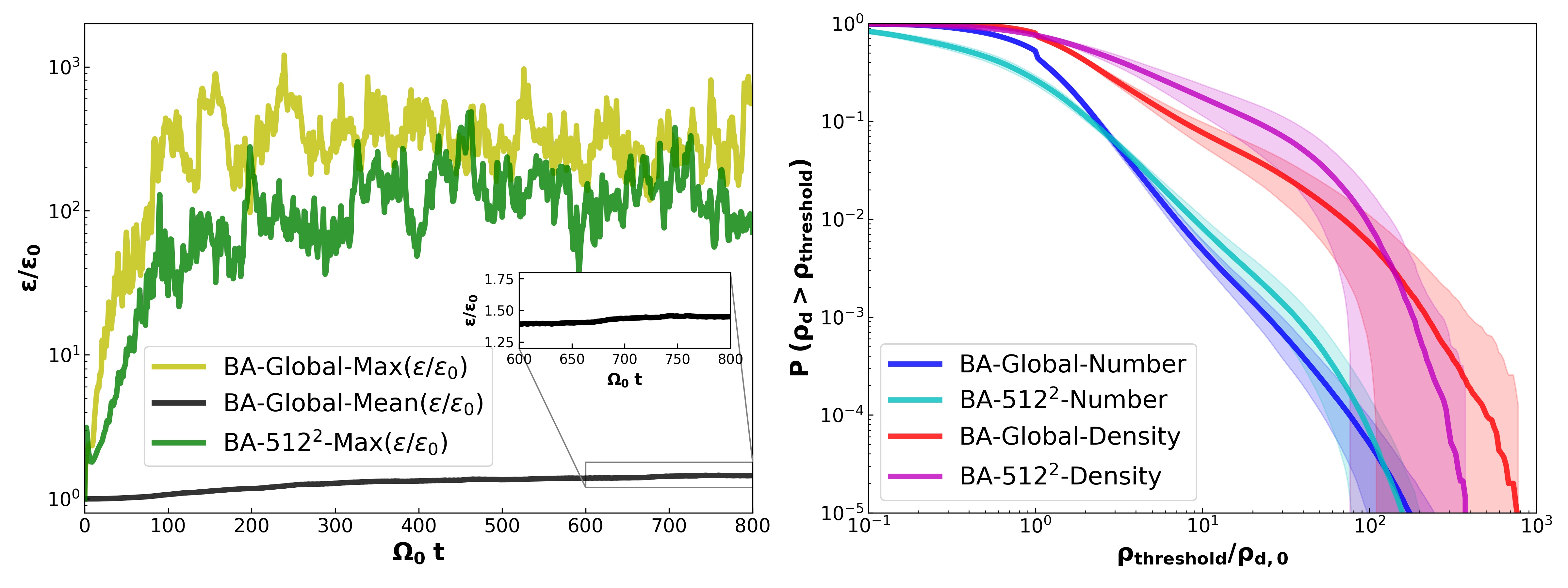}
\caption{Left: The evolution of maximum and mean dust-to-gas ratio $\epsilon$ over time in the global and the local BA-$512^2$. The calculation of $\epsilon$ in the global run are conducted over $r\in[0.4, 2.4]$. Right: Dust CDFs of the global and the local BA-$512^2$, similar to Figure~\ref{fig:cumulative_density}. The dust CDFs of the global run is calculated by the normalized dust density $\rho_\tx{d}r^{0.5}$ on $r\in [0.4, 2.4]$. The color shaded regions are the temporal standard deviations from $600\;\Omega_0^{-1}$ and $800\;\Omega_0^{-1}$. }
\label{fig:Convergence_Global_RunBA}
\end{figure*}

More quantitatively, we have also examined the dust-to-gas ratio $\epsilon$ and the dust CDFs of the global run, and compared them with those from the local BA-$512^2$ shown in Figure~\ref{fig:Convergence_Global_RunBA}. The maximum $\epsilon$ in the global simulation is higher than that in the local simulation by a small margin. Note that due to the reduction of mean radial drift speed in BA test as the SI saturates, and that the SI develops faster in the inner region than the outer region, the mean dust-to-gas ratio $\epsilon$ in the global run increases with time and reaches about $0.3$ (rather than the initial value of $0.2$) at $t = 800\;\Omega_0^{-1}$. This is likely the main cause of the stronger dust clumping found in the global run. On the other hand,
from the dust CDFs, we see that the dust density distribution in the global and local runs converge within the error bars at relatively high densities large $\rho_\tx{threshold}$ ($\gtrsim 5\;\rho_\tx{d}$ both by counting numbers and counting density), suggesting that dust clumping is well captured in both local and global simulations. However, there are
some deviations at relatively small dust density. The cause of this deviation is unknown and may require further investigations beyond the scope of this work, but we speculate it may be related to a combination of higher pressure gradient $\Pi = 0.075$ instead of $0.05$, higher mean $\epsilon$, and the global nature of the simulation.

\subsection{Mesh Refinement}\label{subsec:MR}

In this subsection, we present additional tests to demonstrate the compatibility of our multifluid dust module with static and adaptive mesh refinement (SMR/AMR). Following the convention, here the root mesh is called level 0, and each level of refinement doubles the resolution and is called level 1, 2, etc.

\subsubsection{SMR Run of AB test}\label{subsubsec:RunAB_SMR}

\begin{figure*}[htp]
\centering
\includegraphics[scale=0.6]{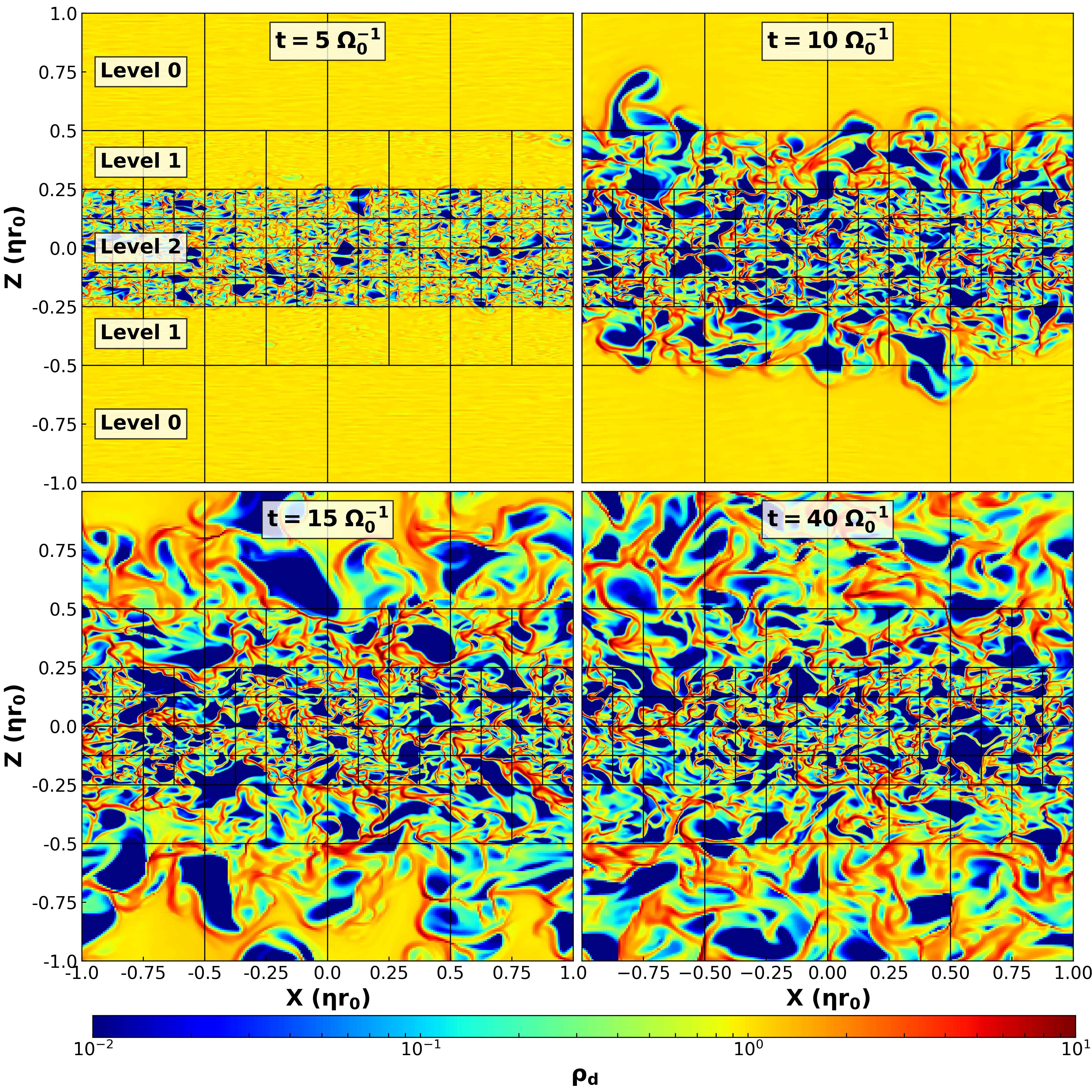}
\caption{Snapshots of dust densities in the SMR test of AB test with two levels of refinement. From top to bottom, from left to right, the panels are at $t = 5\;\Omega_0^{-1}$, $10\;\Omega_0^{-1}$, $15\;\Omega_0^{-1}$ and $40\;\Omega_0^{-1}$. The edges of each meshblock are marked by black solid lines, and meshblocks with different levels are labeled in the top-left panel. Each meshblock contains $64^2$ cells, and there are 8, 16 and 64 meshblocks in levels 0, 1 and 2.
}
\label{fig:SMR_RunAB}
\end{figure*}

We first rerun the AB test of the SI (see Section~\ref{subsubsec:SINonLinear}), but with two levels of SMR. The resolution of the root mesh is $256^2$, and each level of refinement doubles the resolution in the central region along the $z$ direction. Because AB test is sensitive to the amplitude of initial perturbations, we use perturbations ten times smaller ($~0.002\;c_\tx{s}$) than those in uniform runs. The results are shown in Figure~\ref{fig:SMR_RunAB}. As noted earlier, the outcome of AB test depends on resolutions. Indeed, we see that the SI is first developed in the finest level and quickly becomes nonlinear well within one orbital time, while the SI is developed more slowly in coarser meshblocks, and it is not until after about $t \simeq 15\;\Omega_0^{-1}$ that the SI is fully developed in the entire domain. The overall pattern in each refinement level closely resembles those shown in Figure~\ref{fig:Nonlinear_SI} with the same resolution ($256^2$ to $1024^2$), and there are no abrupt features seen along coarse-fine meshblock boundaries, which testifies the compatibility of our multifluid dust module with SMR in shearing box. We have also examined the CDFs of this SMR run, and found that the CDFs at a given level are largely consistent with the CDFs in the corresponding uniform-level runs discussed earlier within the $1\sigma$ uncertainties.

\subsubsection{AMR Test of Kelvin-Helmholtz Instability}\label{subsubsec:KHI}

\begin{figure*}[htp]
\centering
\includegraphics[scale=0.35]{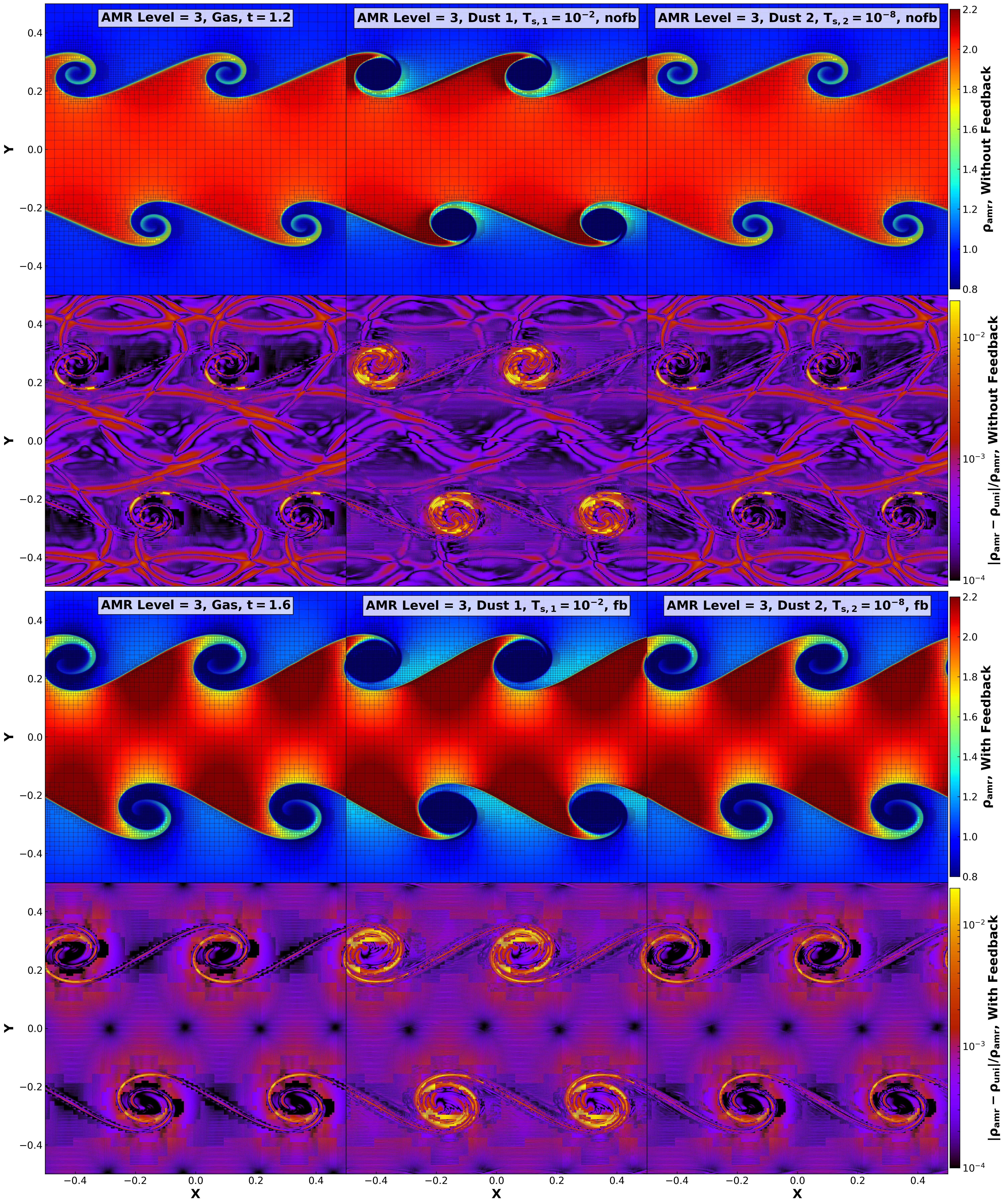}
\caption{Snapshots of the KHI tests with 2 dust species with AMR. The top (bottom) six panels are the cases without (with) dust feedback at $t = 1.2$ ($t = 1.6$). The abbreviations of ``nofb'' and ``fb'' are for the cases without and with feedback, respectively. From left to right, the panels are for gas, dust species 1 and 2. The stopping times of dust are $T_\tx{s,1} = 10^{-2}$ and $T_\tx{s,2} = 10^{-8}$. The first and the third rows are the gas and dust densities from AMR runs with up to 3 refinement levels. The second and the fourth rows are the relative differences between the AMR runs and the uniform grid runs at a resolution matching the finest AMR refinement level of 3. The edges of meshblocks in the AMR runs are also indicated in black solid lines.}
\label{fig:KHI_AMR}
\end{figure*}

We next conduct the standard test problem of the Kelvin-Helmholtz instability (KHI) in Athena++ with AMR, exactly following the problem setup described in Section 3.4.3 of~\cite{Stone2020}, but adding two dust species. The resolution of the root mesh is $256^2$ with each meshblock size being $8^2$, and the refinement condition is determined by
\begin{equation}
  g = \max{(|\pt_x v_{\tx{g},y}|, |\pt_y v_{\tx{g}, x}|, |\pt_x v_{\tx{d},n,y}|, |\pt_y v_{\tx{d},n,x}|)}\ .\\
  \label{eq:condition_KHI}
\end{equation}
which represents the maximum spatial velocity gradients in gas and all dust fluids. Meshblocks with $g > 0.01$ will be refined, and with $g < 0.005$ will be de-refined. We set a maximum of 3 refinement levels. We add two dust species with stopping times $T_\tx{s,1} = 10^{-2}$, $T_\tx{s,2} = 10^{-8}$ but no dust diffusion. We consider the cases without and with dust feedback, and the dust-to-gas mass ratio for each species is set to unity. We use the ``RK2-Implicit'' drag integrator, PLM reconstruction for both gas and dust, the HLLC Riemann solver on gas, and a CFL number of 0.4. For comparison, simulations with a uniform resolution of $2048^2$ (matching the finest level) are also conducted.

Figure~\ref{fig:KHI_AMR} shows the results of our dusty KHI tests with AMR. The first and the third rows show gas and dust density patterns, while the second and the fourth rows show the relative differences from the runs with uniform resolution. We see that the strongly coupled dust with $T_\tx{s,2}=10^{-8}$ shares exactly the same density pattern as gas, as expected. The more marginally coupled dust with $T_\tx{s,1}=10^{-2}$, on the other hand, are depleted in vortex centers, as they are relatively slow in response to the rapid vortical motion to fill in the vortex eyes. We also see that the size of the vortices are larger when feedback is included, as the inertia from more dust loading would require more space for the KHI patterns to roll over.

We find there are no distinguishable differences between the AMR and uniform grid runs, with relative differences of at most a few percent at vortex centers in the first dust species with $T_\tx{s,1} = 10^{-2}$, due in part to the low dust densities in there. The time step in our KHI tests is around $5\times 10^{-5}$ in code units, which is much larger than the stopping time of the second dust species ($T_\tx{s,2} = 10^{-8}$), making the drag interaction highly stiff. The results again testify that our fully-implicit methods are thus accurate and robust in these extremely stiff regimes with AMR.

\section{Summary and Discussion}~\label{sec:summary}

In this paper, we describe the algorithm and implementation of a multifluid dust module in Athena++, together with a suite of benchmark numerical tests. The dust is treated as an arbitrary number of separate pressureless fluids, each interacting with the gas via the aerodynamic gas drag, characterized by a stopping time. Our development features two major advances.

First, we have provided a consistent formulation of dust concentration diffusion. Dust concentration diffusion is commonly implemented as a diffusion term in the dust continuity equation, which mimics the response to background turbulence without explicitly simulating turbulence. This approach has been shown to not conserve angular momentum in disk problems~\citep{Tominaga2019SGI}. We further derive from a Reynolds averaging procedure the proper terms that should be included in the momentum equation to ensure not only proper momentum diffusion flux, but also Galilean invariance. The physically meaningful behavior of dust concentration diffusion including dust feedback is then illustrated from a simple test problem.

Second, we have developed two fully-implicit, second-order accurate drag integrators, which naturally combine with the existing VL2 and RK2 integrators in Athena++ to ensure 2nd-order accuracy in time for the composite system, together with momentum conservation to machine precision. The integrators are stable to highly stiff regimes not only in small dust stopping time, but also in regimes of high dust mass loading. Our code is, to our knowledge, the first to achieve the combination of these features. We have also implemented a number of explicit and semi-implicit drag integrators for non-stiff applications. In addition, we have incorporated frictional heating that can be applied to any of the drag integrators.

The development of the multifluid dust module in Athena++, a higher-order Godunov code, compliments the multifluid dust module in the widely used FARGO3D code~\citep{Benitez2019FARGO3D}, which is Zeus-like~\citep{StoneNorman1992I,StoneNorman1992II}. We conducted a large suite of code tests demonstrating code performance, many in parallel to those done in~\citep{Benitez2019FARGO3D}. In particular, we studied the SI from linear to nonlinear regimes, and the results are generally in good agreement. We anticipate that the aforementioned new features in our implementation represent more benefits, in addition to better shock-capturing capabilities inherent to Godunov codes.

One of the main advantages in our multifluid dust module is its compatibility with many of the existing functionalities and physics modules in Athena++. In particular, our dust fluid module is compatible with static and adaptive mesh refinement, curvilinear coordinate system including cylindrical and spherical coordinates, shearing box and orbital advection, magnetic fields, diffusion physics (viscosity, thermal conduction, non-ideal MHD), etc. The implementation of the multifluid dust module thus enables a wide range of applications involving dust dynamics, particularly related to the study of physics, gas dynamics and observational signatures of protoplanetary disks and planet formation, especially relevant to current and future disk observations by ALMA, James Webb Space Telescope (JWST), Chinese Space Station Telescope (CSST), the Next Generation Very Large Array (ngVLA) and the Square Kilometer Array (SKA). We will also make this module publicly available in the near future to benefit the broader astrophysical community.

There is still substantial room for further extensions of the multifluid dust module, including dust coagulation/fragmentation~\citep{Drazkowska2019}, coupling with non-equilibrium radiative heating and cooling~\citep{KampDullemond2004}, and self-gravity. Additionally, the drag term is exactly the same as coupling among charged and neutral species in weakly ionized plasmas, e.g.,~\cite{O'SullivanDownes2006}, and the coupling term can be extremely stiff in the strong coupling regime. Thus, our code can also be potentially extended to accurately handle weakly ionized plasmas from multifluid to strong coupling regimes. These directions will be considered in future works.

\begin{acknowledgments}
We thank Pablo Ben\'itez-Llambay, Leonardo Krapp, Hui Li, Shengtai Li, Ruobing Dong, Rixin Li, and Shangfei Liu for helpful discussions. This work is supported by the National Key R\&D Program of China No. 2019YFA0405100, and the China Manned Space Project with NO. CMS-CSST-2021-B09.
\end{acknowledgments}

\vspace{5mm}

\software{astropy~\citep{Robitaille2013astropy}, Athena++~\citep{Stone2020}, Mathematica~\citep{Wolfram1991Mathematica}}

\appendix

\section[]{Galilean invariance in momentum diffusion}\label{app:galilean}

In this Appendix, we prove the Galilean invariance of our dust concentration diffusion formulation. In doing so, it would be much easier to recast the dust momentum equation into an Euler-like equation. By applying the dust continuity equation (\ref{eq:dust_con}) and after some algebra, we arrive at
\begin{equation}
\begin{aligned}\label{eq:dust_euler}
\frac{\partial}{\partial t}(\mbs{v}_\tx{d}+\mbs{v}_\tx{d,dif})+
[(\mbs{v}_\tx{d}+\mbs{v}_\tx{d,dif})\cdot\nabla](\mbs{v}_\tx{d}+\mbs{v}_\tx{d,dif})
=\frac{1}{\rho_\tx{d}}\nabla\cdot(\rho_\tx{d}\mbs{v}_\tx{d,dif}\mbs{v}_\tx{d,dif})+\frac{\mbs{v}_\tx{g}-\mbs{v}_\tx{d}}{T_s}\ ,
\end{aligned}
\end{equation}

Let us consider a different frame which moves at a constant speed $\mbs{v}_0$, where physical quantities are denoted with a prime $'$. Obviously, we have $\rho'=\rho$, $\mbs{v}'=\mbs{v}-\mbs{v}_0$, and $\mbs{v}'_\tx{d,dif}=\mbs{v}_\tx{d,dif}$. To prove that the equation is Galilean invariant, it suffices to show that the equations written in the new frame is exactly the same when expressed with primed quantities.

With Equation (\ref{eq:dust_euler}), the proof becomes quite straightforward. We first replace $\mbs{v}_\tx{d}$ by $\mbs{v}'_\tx{d}+\mbs{v}_0$, $\mbs{v}_\tx{g}$ by $\mbs{v}'_\tx{g}+\mbs{v}_0$, $\rho$ by $\rho'$, and $\mbs{v}_\tx{d,dif}$ by $\mbs{v}'_\tx{d,dif}$. We note that the right hand side remains unchanged. Next, note that $\partial/\partial t'=\partial/\partial t+\mbs{v}_0\cdot\nabla$, we can see that terms proportional to $\mbs{v}_0$ all cancel out, thus the form of equation remain exactly the same as (\ref{eq:dust_euler}) except all expressed in primed quantities.

In the above discussion, we emphasize that if drop off the time derivative term on $\rho_\tx{d}\mbs{v}_\tx{d,dif}$, the Euler-like equation would become
\begin{equation}
\begin{aligned}\label{eq:dust_euler2}
\frac{\partial \mbs{v}_\tx{d}}{\partial t}+
[(\mbs{v}_\tx{d}+2\mbs{v}_\tx{d,dif})\cdot\nabla]\mbs{v}_\tx{d}+\frac{\mbs{v}_\tx{d}}{\rho_\tx{d}}\nabla\cdot(\rho_\tx{d}\mbs{v}_\tx{d,dif})
=\frac{\mbs{v}_\tx{g}-\mbs{v}_\tx{d}}{T_s}\ .
\end{aligned}
\end{equation}
which is very different from that of Equation (\ref{eq:dust_euler}). In particular, going through the same procedures, the presence of the third term on the left hand side makes this equation violate the Galilean invariance.

\section{Design of second order fully-implicit drag integrators}~\label{app:imp}

The VL2 implicit integrator is designed from the following format:
\begin{equation}
\begin{aligned}
\mbs{M}^{(n+1)} - \mbs{M}^{(n)}&= h \mbs{f}\left(\mbs{M}^{(n+\frac{1}{2})}, \mbs{W}^{(n+\frac{1}{2})}\right)\\
&=h \mbs{f}\left[\mbs{M}^{(n+1)}- \frac{h}{2}\mbs{f}\left(\mbs{M}^{(n+1)}, \mbs{W}^{(n+\frac{1}{2})}\right), \mbs{W}^{(n+\frac{1}{2})}\right]\\
&=h \mbs{f}\left(\mbs{M}^{(n+1)}, \mbs{W}^{(n+\frac{1}{2})}\right)
- \frac{h^2}{2} \frac{\pt \mbs{f}}{\pt \mbs{M}}|^{(n+\frac{1}{2})}\mbs{f}\left(\mbs{M}^{(n+1)}, \mbs{W}^{(n+\frac{1}{2})}\right)\\
&=h \mbs{f}\left(\mbs{M}^{(n)}, \mbs{W}^{(n+\frac{1}{2})}\right) + h \frac{\pt \mbs{f}}{\pt \mbs{M}}|^{(n)} \left( \mbs{M}^{(n+1)} - \mbs{M}^{(n)} \right)\\
&\quad -\frac{h^2}{2} \frac{\pt \mbs{f}}{\pt \mbs{M}}|^{(n+\frac{1}{2})}\mbs{f}\left(\mbs{M}^{(n)}, \mbs{W}^{(n+\frac{1}{2})}\right)
- \frac{h^2}{2} \frac{\pt \mbs{f}}{\pt \mbs{M}}|^{(n+\frac{1}{2})} \frac{\pt \mbs{f}}{\pt \mbs{M}}|^{(n)}\left( \mbs{M}^{(n+1)} - \mbs{M}^{(n)} \right)\ .
\end{aligned}
\end{equation}
This leads to the integration scheme (\ref{eq:VL2IMPLICIT-2}) and (\ref{eq:LamVL2}).
The quantities at $n+\frac{1}{2}$ time step (denoted by a prime in Section~\ref{subsubsec:2ndIm}) are obtained from the first stage of the algorithm, and any first-order implicit integration suffices (we use the backward Euler method).


The ``RK2-Implicit'' integrator is designed from the following format:
\begin{equation}
\begin{aligned}
  \mbs{M}^{(n+1)} - \mbs{M}^{(n)} &=
  \frac{h}{2} \mbs{f}\left(\mbs{M}^{(n+1)}, \mbs{W}'\right) + \frac{h}{2} \mbs{f}\left(\mbs{M}^{(n)}, \mbs{W}^{(n)}\right) \\
  &= \frac{h}{2} \mbs{f}\left(\mbs{M}^{(n+1)}, \mbs{W}'\right)
  + \frac{h}{2} \mbs{f}\left[\mbs{M}^{(n+1)}- h \mbs{f}\left(\mbs{M}^{(n+1)}, \mbs{W}^{(n)}\right), \mbs{W}^{(n)}\right] \\
  &= \frac{h}{2} \mbs{f}\left(\mbs{M}^{(n+1)}, \mbs{W}'\right) + \frac{h}{2} \left(\mathsf{I} - h \frac{\pt \mbs{f}}{\pt \mbs{M}}\bigg|'\right) \mbs{f}\left(\mbs{M}^{(n+1)}, \mbs{W}^{(n)}\right) \\
  &= \frac{h}{2} \mbs{f}\left(\mbs{M}^{(n)}, \mbs{W}'\right) + \frac{h}{2} \frac{\pt \mbs{f}}{\pt \mbs{M}}\bigg|^{(n)} \left(\mbs{M}^{(n+1)} - \mbs{M}^{(n)}\right)\\
  &\quad+\frac{h}{2} \left(\mathsf{I} - h \frac{\pt \mbs{f}}{\pt \mbs{M}}\bigg|'\right) \mbs{f}\left(\mbs{M}^{(n)}, \mbs{W}^{(n)}\right)
  +\frac{h}{2} \left(\mathsf{I} - h \frac{\pt \mbs{f}}{\pt \mbs{M}}\bigg|'\right) \frac{\pt \mbs{f}}{\pt \mbs{M}}\bigg|^{(n)}\left(\mbs{M}^{(n+1)} - \mbs{M}^{(n)}\right)\ .
\end{aligned}
\end{equation}
where the prime $'$ on $W$ indicate the quantity is evaluated at the end of the first stage (integrating to time step $n+1$ with a first-order implicit scheme, i.e., backward Euler). This leads to the integration scheme (\ref{eq:RK2IMPLICIT-2}) and (\ref{eq:LamRK2}).

\section{Explicit and Semi-implicit Drag Integrators}~\label{app:otherdrag}

When the stopping time of dust is much larger than the numerical time step, $T_\tx{s} \gtrsim \delta t\equiv h$ and there is no strong dust mass loading, the drag is in non-stiff regime. Here we have also implemented a number of standard explicit and semi-implicit drag integrators in Athena++. We omit the trivial implementation of the forward Euler method (RK1), and 2nd-order methods are described below. For all methods, the energy equation is updated in each stage in a way analogous to Equations (\ref{eq:eng1}) to (\ref{eq:eng3}), which we again omit here.

\subsection{Second Order Explicit Methods}\label{subsubsec:2ndEx}

Here we document the two explicit integrators following the VL2 and RK2 integrators in Athena++, termed ``\txbf{VL2-Explicit}'' and ``\txbf{RK2-Explicit}'' in this paper. Note that the explicit integrators usually requires the time step $h<T_\tx{s}$ for all dust species. The momentum update is as follows.

\txbf{VL2-Explicit}:
We first update the system for half a time step, followed by a full update using the midpoint values:
\begin{equation}
\label{eq:VL2EXPLICIT}
\begin{aligned}
  \mbs{M}^{(n+\frac{1}{2})} &= \mbs{M}^{(n)}+\frac{h}{2} \mbs{f}\left(\mbs{M}^{(n)}, \mbs{W}^{(n)}\right)\ ,\\
  \mbs{M}^{(n+1)}   &= \mbs{M}^{(n)}+h \mbs{f}\left(\mbs{M}^{(n+\frac{1}{2})}, \mbs{W}^{(n+\frac{1}{2})}\right)\ .\\
\end{aligned}
\end{equation}

\txbf{RK2-Explicit}: it first provides an estimate after a time step $h$ denoted by ${'}$, followed by a correction:
\begin{equation}
\label{eq:RK2EXPLICIT-1}
\begin{aligned}
  \mbs{M}' &= \mbs{M}^{(n)}+h \mbs{f}\left(\mbs{M}^{(n)}, \mbs{W}^{(n)}\right)\ ,\\
  \mbs{M}^{(n+1)}&= \frac{1}{2} \left(\mbs{M}^{(n)} + \mbs{M}'\right) + \frac{1}{2} h \mbs{f}\left(\mbs{M}', \mbs{W}'\right)\ .
\end{aligned}
\end{equation}

\subsection{Second Order Semi-implicit Methods}\label{subsubsec:semiIm}

Here we present two semi-implicit methods, which are more robust than the explicit methods.

\txbf{Trapezoid (Crank-Nicholson Method)}

The trapezoid method is derived from

\begin{equation}
\label{eq:Trapezoid-0}
\begin{aligned}
  \mbs{M}^{(n+1)}&=\mbs{M}^{(n)}+\frac{1}{2} \left[h \mbs{f}\left(\mbs{M}^{(n)}, \mbs{W}^{(n)}\right) + h \mbs{f}\left(\mbs{M}^{(n+1)}, \mbs{W}^{(n)}\right)\right],\\
  &=\mbs{M}^{(n)}+h \mbs{f}\left(\mbs{M}^{(n)}, \mbs{W}^{(n)}\right) + \frac{h}{2} \frac{\partial \mbs{f}}{\partial \mbs{M}}|^{(n)} \left(\mbs{M}^{(n+1)} - \mbs{M}^{(n)}\right),\\
  \Rightarrow \mbs{M}^{(n+1)} &= \mbs{M}^{(n)} + \left(\mathsf{I} - \frac{h}{2} \frac{\pt \mbs{f}}{\pt \mbs{M}}|^{(n)}\right)^{-1} h \mbs{f} \left(\mbs{M}^{(n)}, \mbs{W}^{(n)}\right).\\
\end{aligned}
\end{equation}

The first stage $'$ of ``Trapezoid'' is updated by backward Euler method with $h$, so as to be compatible with  ``RK2-Explicit''.

\txbf{Trapezoid Backward Differentiation Formula 2 (TrBDF2)}

In the TrBDF2, the momentum at the middle stage $n+\frac{1}{2}$ is calculated by the trapezoid method with time step $h/2$, so as to be compatible with ``VL2-Explicit''. Then $\mbs{M}^{(n+1)}$ is updated by backward Differentiation Formula 2 (BDF2) method at the stage $n+1$.

\begin{equation}
\label{eq:TR-BDF2-0}
\begin{aligned}
  \mbs{M}^{(n+\frac{1}{2})} &= \mbs{M}^{(n)} + \left(\mathsf{I} - \frac{h}{4} \frac{\pt \mbs{f}}{\pt \mbs{M}}|^{(n)}\right)^{-1} \frac{h}{2} \mbs{f} \left(\mbs{M}^{(n)}, \mbs{W}^{(n)}\right),\\
  \mbs{M}^{(n+1)}&=\frac{4}{3} \mbs{M}^{(n+\frac{1}{2})}- \frac{1}{3} \mbs{M}^{(n)} + \frac{1}{3} \left(\mathsf{I} - h \frac{\partial \mbs{f}}{\partial \mbs{M}}|^{(n)}\right)^{-1} h \mbs{f}\left(\mbs{M}^{(n)}, \mbs{V}^{(n)}\right).
\end{aligned}
\end{equation}

\section{Performance}\label{app:performance}

\begin{figure*}[htp]
\centering
\includegraphics[scale=0.50]{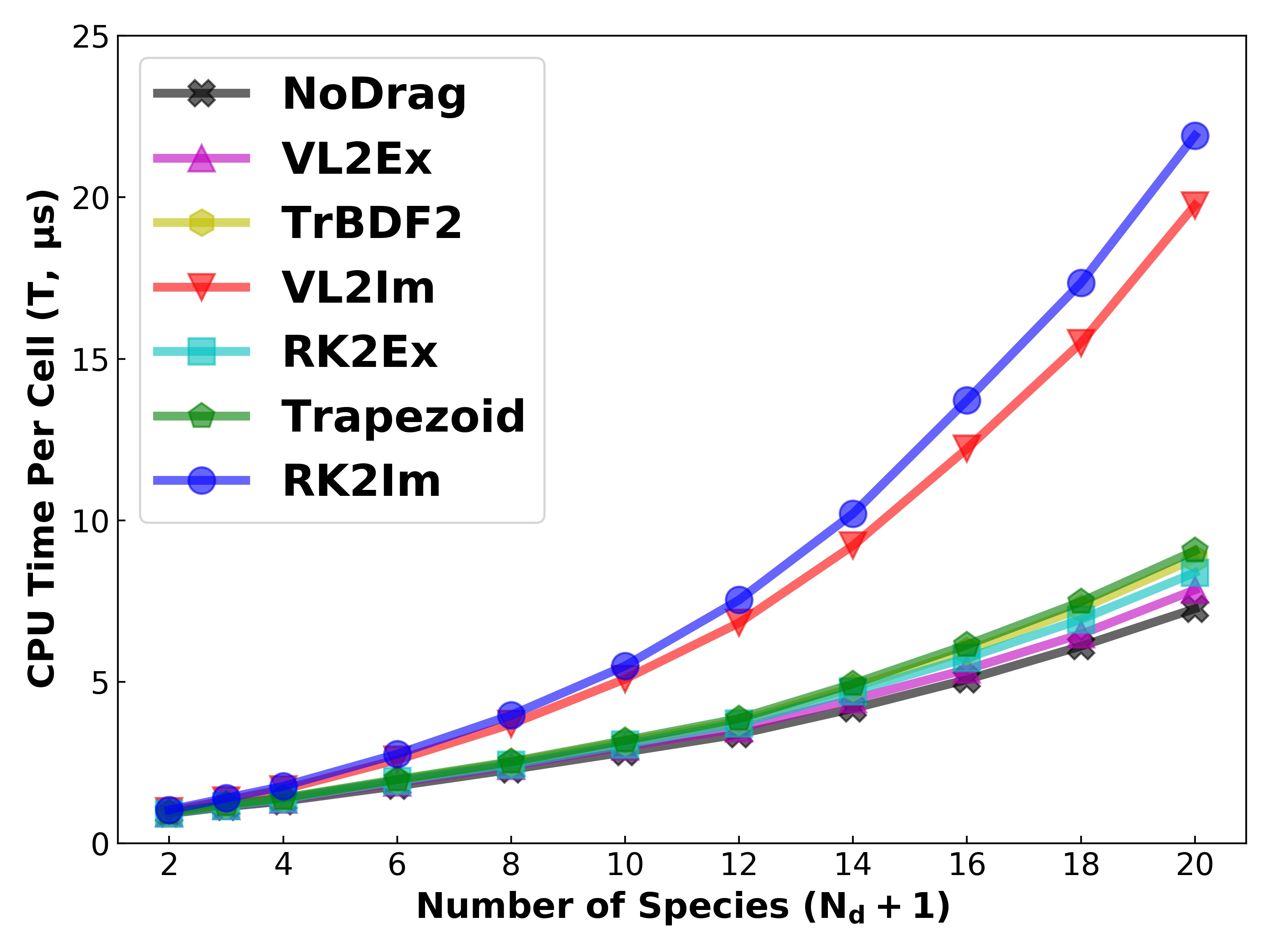}
\caption{Code performance measured by CPU time per cell in micron second, shown as a function of total number of species (gas and $\mbf{N_{d}}$ species of dust).}
\label{fig:performance}
\end{figure*}

The two higher-order fully-implicit drag integrators VL2-Implicit and RK2-Implicit involve solving the inverse of an $(N_d+1)\times(N_d+1)$ matrix. The cost of matrix inversion is $O[(N_d+1)^3]$. Moreover, it also takes $O[(N_d+1)^3]$ to handle matrix multiplications, such as in computing $\mbs\Lambda$ in Equations (\ref{eq:LamVL2}) and (\ref{eq:LamRK2}). Such operations would make the calculations increasingly expensive as $N_d$ increases, and can become a bottleneck at sufficiently large $N_d$.

Here we measure code performance as a function of $N_d$ from the NSH equilibrium test in shearing box. The test is run on a Intel Xeon Gold 6132 CPU with 28 cores. We use all the cores with 28 meshblocks, so that we occupy the entire CPU (and hence its cache) to mimic more realistic situation in large-scale simulations (note that communications in Athena++ are mostly hidden thanks its use of tasklist and performance is more sensitive to catch use). We use the HLLE Riemann solver for gas and the PLM reconstruction for both gas and dust. We measure the performance in terms of the time spent to update a single cell by an individual core. The results for different drag integrators are shown in Figure~\ref{fig:performance}, as a function of total number of species ($N_d+1$), which are further compared to results with gas drag turned off.

For explicit and semi-implicit integrators, we see that the drag integrators add to very limited computational cost relative to the no-drag case. In particular, the semi-implicit integrators that involve two matrix inversion operations remain computationally efficient thanks to the fast analytical solver~\citep{KrappBenitez2020RNAAS} that reduces the cost to $O(N_d+1)$. The total cost increases linearly with $N_d$ for $N_d\lesssim12$, but gets slightly nonlinear at larger $N_d$. We speculate it is likely due to heavier memory use that reduces cache performance.

For the two fully implicit solvers, we manage to improve the performance by using the fast matrix inversion at the first integration stage, yet the more complex matrix computation and inversion at the second stage substantially increases the computational cost. This cost increases non-linearly with $N_d$. It is relatively negligible for $N_d\lesssim5$, and remains to be minor compared to the rest of the dust integrator for $N_d\lesssim10$, but becomes rather significant for larger $N_d$.

\section{Solutions to the Collision Tests}~\label{app:ana_collision}

The mutual drags between gas and $n$ dust species can be written in the matrix form:
\begin{small}
\begin{equation}
\centering
\begin{aligned}
  \frac{\pt \mbs{M}}{\pt t} = \mbs{A} \mbs{M} =
\begin{bmatrix}
-\sum^n_i \epsilon_i \alpha_i & \alpha_1  & \alpha_2  & \cdots & \alpha_n  \\
\epsilon_1 \alpha_1           & -\alpha_1 & 0         & \cdots & 0         \\
\epsilon_2 \alpha_2           & 0         & -\alpha_2 & \cdots & 0         \\
\vdots                        & \vdots    & \vdots    & \ddots & \vdots    \\
\epsilon_n \alpha_n           & 0         & 0         & \cdots & -\alpha_n \\
\end{bmatrix}
\begin{bmatrix}
\mbs{M}_\tx{g}\\
\mbs{M}_{\tx{d},1}\\
\mbs{M}_{\tx{d},2}\\
\vdots \\
\mbs{M}_{\tx{d},n}\\
\end{bmatrix} \ .
\end{aligned}
\end{equation}
\end{small}
The analytic solution of the momentum vector must take the form
$\mbs{M(t)} = \sum_0^{n}\mbs{c}_i \exp{(\lambda_i\;t)}$, where $\mbs{c}_i$ are the coefficient for each momentum component determined by initial condition and $\lambda_i$ are the eigenvalue. The key is to solve the eigensystem $\mbs{A} \mbs{M}=\lambda \mbs{M}$, yielding eigenvalues $\lambda_i$ and eigenvectors $\mbs{M}_i$. The coefficients $\mbs{c}_i$ are obtained by decomposing the initial condition into the eigenvectors.

Because of momentum conservation, i.e., the drag force acted on the gas equals to the sum of drag forces acted on dust, the matrix $A$ has an eigenvalue $\lambda_0=0$ corresponding to bulk motion, and hence the coefficient $c_0 = v_\tx{COM}$ is the velocity of center of mass (COM). When multiple dust species share the same stopping time, the eigensystem can be greatly simplified (see Table 1 of~\cite{Benitez2019FARGO3D}), but this is no longer true in the general case. We use Mathematica~\citep{Wolfram1991Mathematica} to calculate the eigenvalues and the rest of the coefficients in Table~\ref{tab:collision}.

\section{Additional Numerical Tests}

In this Appendix, we present additional code tests that largely reproduce the dusty sound wave test and the dusty shock test in \cite{Benitez2019FARGO3D} to demonstrate our code performance.

\subsection{Dusty Sound Wave}~\label{app:dustysoundwave}

\begin{figure*}[htp]
\centering
\includegraphics[scale=0.4]{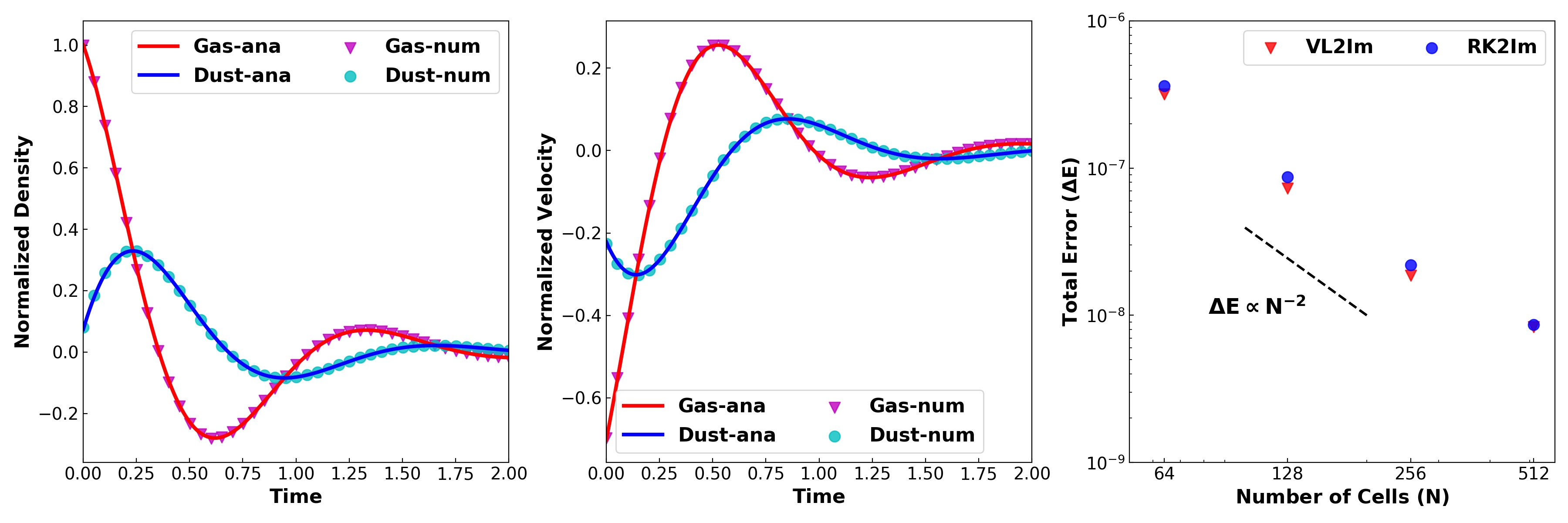}
\caption{Results from the dusty sound wave test. The left and the middle panels are the time evolution of normalized density and normalized velocities of gas and dust located at $x=0$, to be compared with Figure 3 of~\cite{Benitez2019FARGO3D}. The solid lines represent the analytical solutions, and the triangle and circle markers are the numerical results from test runs with 256 cells. The right panel demonstrates the numerical convergence, where we calculate the mean L1-error after one wave period, and show how it varies with the number of grid cells using the ``VL2-Implicit'' and ``RK2-Implicit'' drag integrators.}
\label{fig:dustywave}
\end{figure*}

To demonstrate that our multifluid dust module achieves second order accuracy when combined with the hydrodynamic solver, we conduct the dusty sound wave test by exactly following 
Section 3.2 of~\cite{Benitez2019FARGO3D}, originally proposed by~\cite{LaibePrice2011,LaibePrice2012SPH1}. We use the PLM spatial reconstruction, isothermal equation of state with the HLLE gas Riemann solver, and consider both the ``VL2-Implicit'' and ``RK2-Implict'' drag integrators. The tests are conducted in 1D, starting from a resolution of $N=64$ cells, and we double the resolution at a time until reaching a resolution of $N=512$ cells. We have conducted simulations for both single-species and multi-species cases, and found excellent agreement with analytical theory. For brevity, we show in Figure \ref{fig:dustywave} the time evolution of the normalized dust and gas velocities for the single species dust case, showing that our numerical solution perfectly matches the analytical solution. Moreover, we measure the root mean square of the L1 norms $[\sum_n(\sum|U_n-U_{n,{\rm ana}}|/N)^2]^{1/2}$ after one wave period, similar to the approach in the linear wave test in Athena++ \cite{Stone2020}, where $U_n$ and $U_{n,{\rm ana}}$ are the numerical and analytical solutions of the $n$-th variable. The variables include gas density and velocity, and dust density and velocity, all in normalized units. We see in the right panel of Figure \ref{fig:dustywave} that our code clearly achieves second order convergence.


\subsection{Dusty Shock}~\label{app:dustyshock}

\begin{figure*}[htp]
\centering
\includegraphics[scale=0.5]{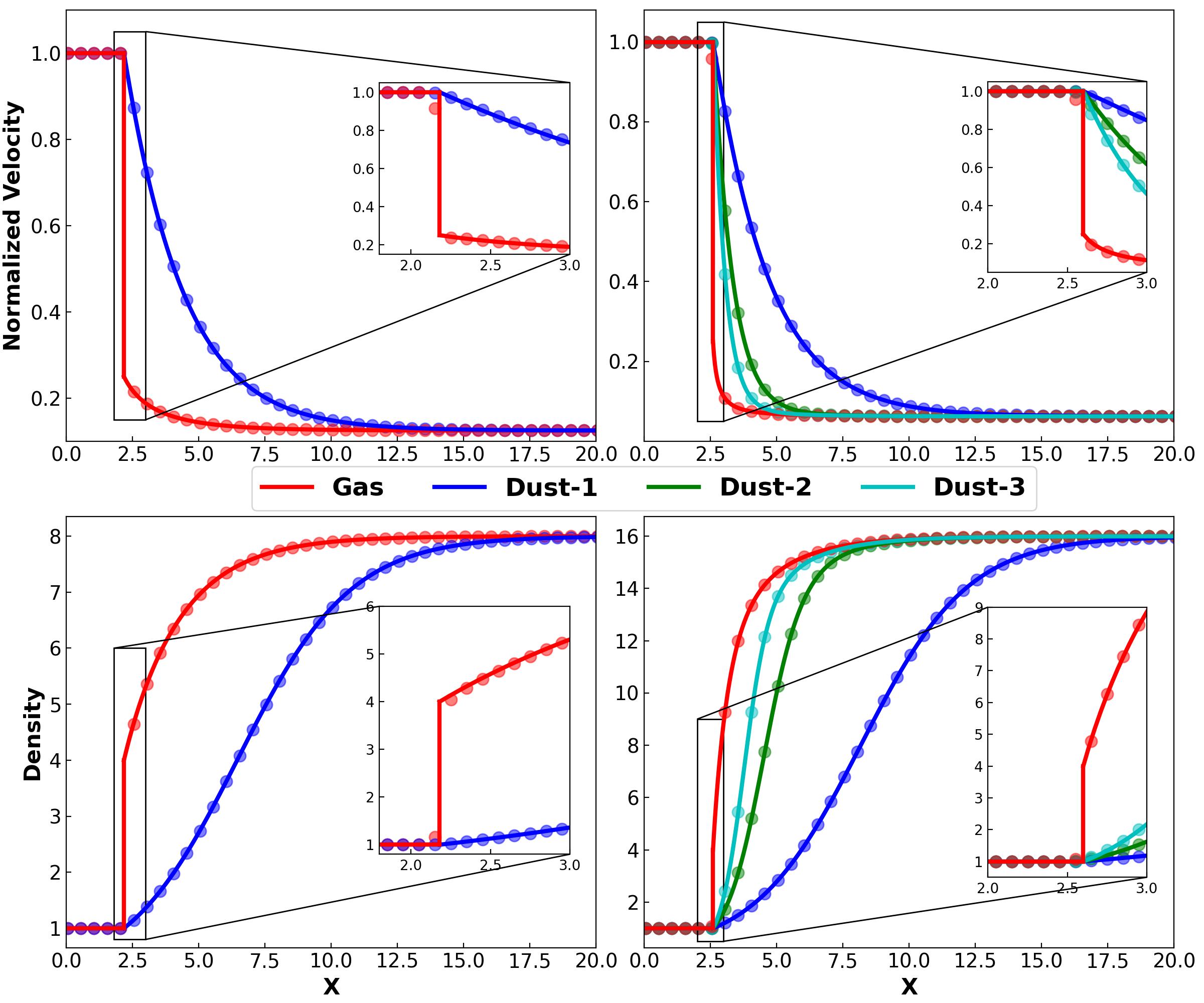}
\caption{Normalized velocities (top) and densities (bottom) of dusty wave tests with 1 (left) and 3 (right) dust species, to be directly compared with Figure 5 of \cite{Benitez2019FARGO3D}. The gas profile is shown in red, while the other colors correspond to the dust species, with the insets showing the zoomed in profiles across the shock. Analytical solutions are shown in solid lines, while numerical results are shown in filled circles.}
\label{fig:dustyshock}
\end{figure*}

Being a Godunov code, Athena++ has excellent shock capturing properties that we demonstrate using the generalized dusty shock test presented in Section 3.3 of~\cite{Benitez2019FARGO3D}, which is generalized from ~\cite{LehmannWardle2018}. We follow the same procedures and adopt identical parameters as in \cite{Benitez2019FARGO3D} to conduct two simulations with one and three dust species (two and four species in total) on 400 grid points. PLM spatial reconstruction and ``VL2-Implicit'' drag integrator are used for these tests. The results are shown in Figure~\ref{fig:dustyshock}, which is to be compared with Figure 5 of \cite{Benitez2019FARGO3D} side by side. Note that the shocks are in steady state thus we focus on the overall shock profile rather than the specific shock locations. Because of dust drag, the dust profile near the shock is connected by a precursor that is accurately reproduced, similar to that in FARGO3D. On the other hand, Athena++ very captures the discontinuity in the gas within neighboring cells, as opposed to $\sim4$ cells in FARGO3D.

\bibliographystyle{aasjournal}
\bibliography{references}{}

\end{document}